\documentclass[preprint,12pt]{elsarticle}

\usepackage[utf8]{inputenc}

\usepackage{amssymb}
\usepackage{amsmath}
\usepackage{booktabs}
\usepackage{tabularx}
\usepackage{multirow}
\usepackage[T1]{fontenc}

\usepackage[version=4]{mhchem}
\usepackage{enumitem,cleveref}
\usepackage{siunitx}
\DeclareSIUnit\angstrom{\text {Å}}
\usepackage{tabularx}
\usepackage{xcolor}
\usepackage{float}
\newfloat{listfloat}{htbp}{lop}
\floatname{listfloat}{Protocol}

\graphicspath{{figures/}}
\RequirePackage{tikz}
\usetikzlibrary{arrows,positioning,calc,shapes.geometric}

\tikzstyle{labeled}=[execute at begin node=$\scriptstyle,
   execute at end node=$]
\RequirePackage{graphicx}

\journal{Computational Materials Science}

\begin{document}

\begin{frontmatter}



\title{Phonon density of states of silica (\ce{SiO2}) nanopore via molecular dynamics simulations.}


\author[ANSTO]{Pablo Galaviz} 
\author[ANSTO]{Dehong Yu}
\author[ANSTO]{Nicolas de Souza}
\author[IbarakiUni]{Sho Kimura}
\author[IbarakiUni]{Yoshitomo Kojima}
\author[IbarakiUni,RECAS]{Seiji Mori}
\author[IbarakiUni]{Akira Yamaguchi}

\affiliation[ANSTO]{organization={Australian Centre for Neutron Scattering, Australian Nuclear Science and Technology Organisation},
            addressline={New Illawarra Road}, 
            city={Lucas Heights},
            postcode={2234}, 
            state={New South Wales},
            country={Australia}}
\affiliation[IbarakiUni]{organization={Institute of Quantum Beam Science, Graduate School of Science and Engineering, Ibaraki University},
            city={Mito},
            state={Ibaraki 310-8512},
            country={Japan}}
\affiliation[RECAS]{organization={Research and Education Center for Atomic Sciences, Ibaraki University},
            city={Tokai},
            state={Ibaraki 319-1106},
            country={Japan}}


\begin{abstract}
This study presents a comprehensive computational investigation of the vibration density of states (VDOS) of a silica nanopore, systematically evaluating a range of force fields against inelastic neutron scattering results. We analyze the influence of temperature, crustal structure, and surface-adsorbed water molecules on the nanopore's structural and dynamic properties. 
We performed classical molecular dynamics simulations of nanopore and bulk silica, and used density functional theory (DFT) calculations for bulk silica for comparison.
The resulting VDOS shows relatively good agreement with DFT and experimental data. The temperature has a relatively low effect on the dry nanopore. The inclusion of \ce{H2O} molecules significantly affects the VDOS. The low energy modes are dominated by \ce{H2O} VDOS and increase with loading. 
This work is an essential step towards characterizing silica nanopores via molecular dynamics and provides a valuable reference for experimental comparison with X-ray and neutron scattering VDOS results. 
\end{abstract}

\begin{graphicalabstract}
\centering
\includegraphics[width=\textwidth]{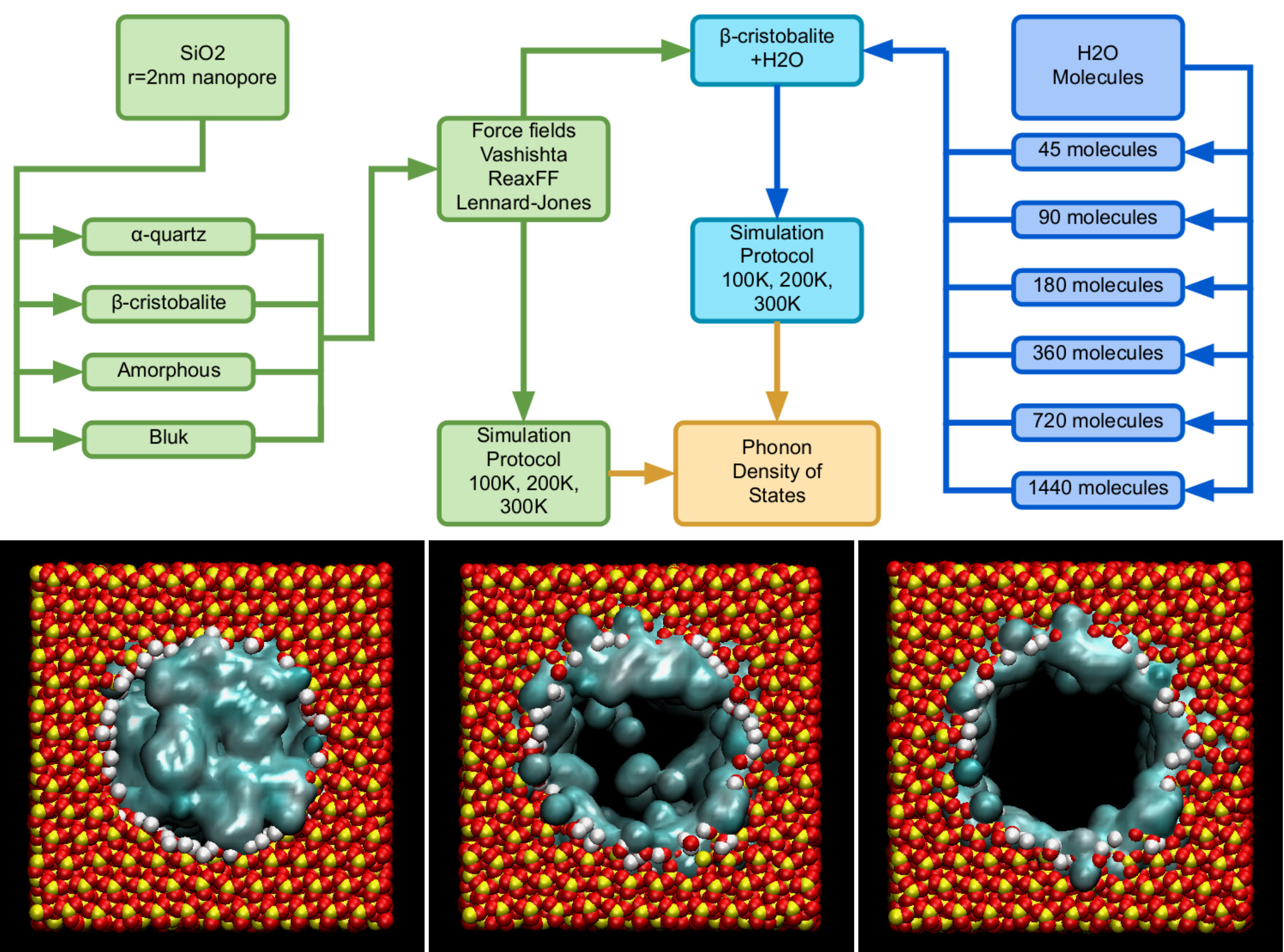}
\end{graphicalabstract}


\begin{highlights}
\item The vibrational density of states in \ce{SiO2} nanopores was studied using molecular dynamics simulations.  

\item Classical force fields are comparable with neutron scattering and density functional theory results.

\item The vibrational density of states significantly depends on temperature between \qty{100}{K} and \qty{400}{K} when loaded with \ce{H2O}.

\item At low temperature, quadratic scaling is observed in the total density of states in the low-energy range.

\item At high temperature, almost linear scaling is observed in the total density of states in the low-energy range.
\end{highlights}


\end{frontmatter}



\section{Introduction}
\label{sec: introduction}

Nanopores are small cavities in materials, typically with diameters ranging from approximately \qty{0.5}{nm} to \qty{100}{nm} nanometers. Nanopores can contain individual molecules such as ions, nucleic acids, or proteins. Nanopores are also nanoscale openings or channels that allow the passage of ions and molecules. They can serve as sensing devices when a voltage is applied across the material. In that instance, ions flow steadily through the nanopore. The insertion of a molecule temporarily impedes ion flow, leading to detectable changes in current amplitude and duration. These transient events underpin nanopore sensing, providing detailed information about a molecule’s size, shape, charge, and conformation.

Nanopores are classified into three categories: biological, solid-state, and hybrid \cite{BhaJawAli2021}. Some examples of biological nanopores include $\alpha$-Hemolysin, MspA, and bacteriophage Phi29. These nanopores have scales between \qty{1}{nm} and \qty{3.6}{nm} diameter. These kinds of nanopores have applications in DNA and RNA sequencing, protein analysis and post-translational modification detection, medical diagnostics and clinical applications, and chemical and environmental sensing \cite{HaqLiWu2013, YinHuZha2022, ZheZhoDin2023, LiuLiTan2023}. Solid-state nanopores are fabricated from synthetic materials like \ce{SiN}, \ce{Al2O3}, \ce{Si3N4}, \ce{SiO2}, \ce{MoS2}, and  graphene using nanofabrication techniques. Other applications of solid-state nanopores beyond sensing technology are photovoltaic and photothermal conversion, thermoelectric and electrokinetic energy conversion, energy storage, separation and filtration, catalysis and gas purification \cite{CheCheXie2025, GarHolLiu2016, LiLiRui2024, PanXuZha2024}. 

Silica (\ce{SiO2}) nanopores have applications across diverse disciplines, including biomedical applications, the fabrication of energy storage materials, nanoelectronic templates, and filtration \cite{XuLiShi2023, XiaLiuKan2023, HeTsuZho2021}. Simulation techniques such as classical density functional theory \cite{YanSuLia2020} and multiphysics methods \cite{GubBalDi2022} have been used to study \ce{SiO2} nanopores. Previous research on molecular dynamics simulations of \ce{SiO2} nanopores includes water vapor outgassing \cite{KimFriNed2015}, the behaviour of \ce{CO2} confined molecules \cite{MohLiuGad2021, XuLiaZha2025}, \ce{CO2} adsorption mechanisms \cite{ShiZhaSun2025}, nanopore surface modelling \cite{FouHanThe2022}, and reverse osmosis \cite{WilWhiBou2022}.

The phonon density-of-states calculation for bulk and thin-film \ce{SiO2} using density functional theory has been previously investigated \cite{ZhuZheCao2018, TsaKacSto2024}. To our knowledge, there are no previous phonon density-of-states simulations of \ce{SiO2} nanopores using molecular dynamics. In this article, we characterize the phonon density of states of \ce{SiO2} nanopores and thoroughly examine several factors that influence it, including temperature, force field, and the addition of surface water. Our results make an important contribution to the study of silica nanopore, providing a valuable reference for experimental comparisons and setting a precedent for simulating other similar nanomaterials. There is room for further exploration, such as the confinement of organic molecules and peptides.

The manuscript is organized as follows. Section \ref{sec: methods} describes the numerical setup and techniques. Section \ref{sec: results} presents the results, including dependencies on the force field, nanopore vs bulk comparison, temperature dependency, as well as the effects of surface-adsorbed water. The appendices provide details on the annealing tests and heating procedure test. Finally, conclusions and discussions are given in Section \ref{sec: conclusions}.

\subsection{Notation and abbreviations}
\label{subsec: notation}

We used two molecular dynamics (MD) software. The \textsc{Large-scale Atomic/Molecular Massively Parallel Simulator} (LAMMPS)\cite{LAMMPS2022} and GROMACS \cite{GROMACS2020}. We calculated the phonon density of states (PDOS) and mean square displacement (MSD) with the \textsc{Molecular Dynamics Analysis for Neutron Scattering Experiments} (MDANSE) \cite{MDANSEref}. When discussing noncrystalline structures, we will refer to vibrational density of states (VDOS) rather than PDOS. We performed density functional theory (DFT) calculations using the \textsc{Quantum Espresso} (QE)\cite{GiaBarBon2009, GiaAndBruBunBuo2017, GiaBasBon2020} and the Vienna Ab initio Simulation Package (VASP) \cite{VASP1, VASP2, VASP3, VASP4, VASP5, VASP6}. The DFT phonon calculation was performed using the finite displacement method (FDM) with the help of \textsc{Phonopy} \cite{Phonopy1,Phonopy2}. The PDOS analysis was done with \textsc{Euphonic} \cite{FaiJacVon2022}. We used isothermal–isobaric (NPT) and canonical (NVT) ensembles. We compared our results with time-of-flight inelastic neutron scattering (TOF-INS). 

\section{Methods}
\label{sec: methods}

\begin{figure}[!ht]
\centering
    \includegraphics[width=\textwidth]{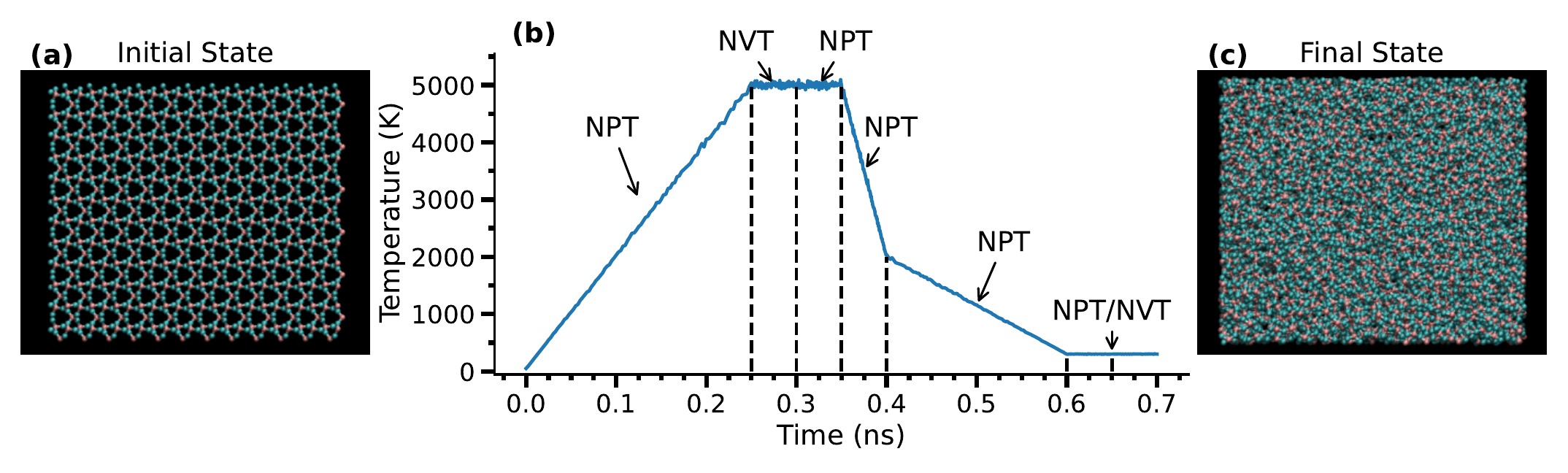}
\caption{Anneling procedure. Each stage shows the thermodynamical thermostat employed} \label{fig: annealing}
\end{figure}

We employ two methods for generating silica nanopores. For GROMACS simulations, we use the Python PoreMS library version 0.2.5 cite{PoreMS}. For LAMMPS simulations, the initial structure was retrieved from the Crystallography Open Database \cite{GraChaDow09}. We use an $\alpha-$quartz structure with ID 1526860 \cite{NeiDol94}. Using a custom Python script, we generate a silica supercell of a given size. From this initial structure, we produced an amorphous silica block by an annealing procedure. For that purpose, we used the following simulation protocol:  
\begin{listfloat}
  \caption{Annealing.}
  \label{protocol: annealing}
\begin{enumerate} 
    \item Initial atom relaxation with tolerance \qty{1000}{kJ.mol^{-1}.nm^{-1}}.
    \item NPT heating from \qty{50}{K} to \qty{5000}{K} for \qty{0.25}{ns} with external pressure increasing from \qty{1}{bar} to \qty{100}{bar}.
    \item NVT equilibration at \qty{5000}{K} during \qty{0.05}{ns}.
    \item NPT equilibration at \qty{5000}{K} with external pressure at \qty{100}{bar} during \qty{0.05}{ns}. 
    \item NPT cooling from \qty{5000}{K} to \qty{2000}{K} for \qty{0.05}{ns} with external pressure decreasing from \qty{100}{bar} to \qty{10}{bar}.
    \item NPT cooling from \qty{2000}{K} to \qty{300}{K} for \qty{0.2}{ns} with external pressure decreasing from \qty{10}{bar} to \qty{1}{bar}.
    \item NPT equilibration at \qty{300}{K} with \qty{1}{bar} external pressure during \qty{0.05}{ns}.
    \item NVT equilibration at \qty{300}{K} during \qty{0.05}{ns}. 
\end{enumerate}
\end{listfloat}
Figure \ref{fig: annealing} shows the temperature and stage of the annealing procedure as a function of time. Using the resulting amorphous block, we created a pore by removing the atoms within an embedded cylinder of a given radius. We coated the interior of the pore with silanol functional groups of type \(Q^2\) [\ce{Si(OH)2}] and \(Q^3\) [\ce{SiOH}]. We performed nanopore simulations with and without silanol groups added. 

We simulated a silica porous of size $r=\qty{2}{nm}$. The numerical domain was a cube of side $L=\qty{6}{nm}$. We imposed periodic boundary conditions. The distance between the pores' images was $\qty{4}{nm}$. We added a given number of \ce{H2O} molecules, randomly distributed within the pore. We tested several force fields. For LAMMPS, we use a \texttt{reaxff} potential \cite{NaoNumJeo2023}, and a Vashishta \cite{ff_Vashishta}. For GROMACS, we used the potential from \cite{KraRybHol2021, AbrIgl2019, GulTho2006} and a TIP4P2005 flexible water model \cite{TIP4P2005f}. We also tested nonflexible water models TIP3P and SPC. However, we found that a flexible water model performs better for vibrational mode calculations. We simulated dry porous and wet porous with 90, 180, 515, 720, and 1440 \ce{H2O} molecules. 

We performed production runs at several temperatures, initially heating the system and progressively cooling it using the following simulation protocol:  
\begin{listfloat}
  \caption{Quick heating and equilibration.}
  \label{protocol: quick}
\begin{enumerate} 
    \item  Initial atom relaxation with tolerance \qty{1000}{kJ.mol^{-1}.nm^{-1}}.
    \item  Configuration starts at $T_0=\qty{10}{K}$, $i=0$, generating a Gaussian velocity distribution that neutralizes total angular and linear momentum.
    \item  $NPT$ heating from $T_i$ to $T_{max}$ for \qty{20}{ps}. 
    \item  $NVT$ equilibration for \qty{20}{ps}. \label{itm: p1 restart}
    \item  $NVT$ phonon sampling for \qty{50}{ps}. 
    \item  If $T_i\leq T_{min}$ then the simulation finishes. 
    \item  Set $T_{i+1} \rightarrow T_i-\Delta T$.
    \item  $NPT$ cooling from $T_i$ to $T_{i+1}$ for \qty{20}{ps}, set $i \rightarrow i+1$ then continue to \ref{itm: p1 restart}
\end{enumerate}
\end{listfloat}

\begin{figure}[!ht]
\centering
    \includegraphics[width=\textwidth]{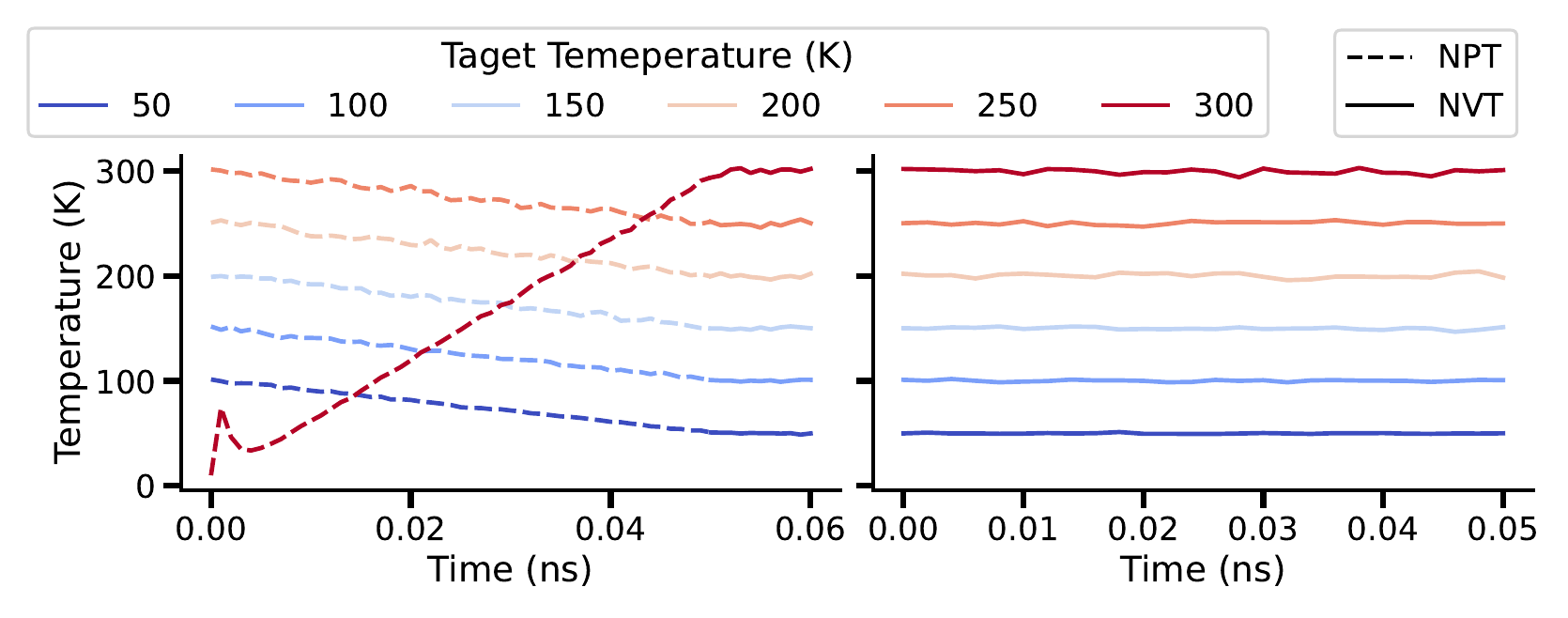}
\caption{Quick simulation protocol.} \label{fig: protocol}
\end{figure}

Here $T_i=\qty{10}{K}$, and $T_{min}=\qty{100}{K},T_{max}=\qty{300}{K}$ and $\Delta T=\qty{50}{K}$. Figure \ref{fig: protocol} shows a typical output of the temperature and the different ensemble stages. We tested an alternative protocol that progressively heats the system. However, we found that the cooling procedure produces an experimentally consistent result with \ce{H2O} molecules properly attached to the pore's surface (see \ref{app: heating_procedure_test}). 

For a more detailed analysis, we performed a long heating and equilibration simulation. In this case, we use the following protocol:
\begin{listfloat}
  \caption{Long heating and equilibration.}
  \label{protocol: long}
\begin{enumerate} 
    \item  Initial atom relaxation with tolerance \qty{1000}{kJ.mol^{-1}.nm^{-1}}.
    \item  Configuration starts at $T_0=\qty{10}{K}$, $i=0$, generating a Gaussian velocity distribution that neutralizes total angular and linear momentum.
    \item  $NPT$ heating from $T_i$ to $T_{i+1}$ for \qty{1}{ns}. \label{itm: p2 restart}
    \item  $NVT$ equilibration for \qty{10}{ns}. 
    \item  $NVT$ phonon sampling for \qty{50}{ps}. 
    \item  If $T_i<T_{max}$, set $T_{i+1} \rightarrow T_i+\Delta T$, and then $i\rightarrow i+1$. Continue to \ref{itm: p2 restart}.
\end{enumerate}
\end{listfloat}

Here $T_1=\qty{100}{K}, T_{max}=\qty{400}{K}$ and $\Delta T=\qty{50}{K}$.

\subsection{Force fields}
\label{sec: forcefields}

We tested three force fields, described in Table \ref{tab: forcefields}.
\begin{table}[h]
    \centering
    \begin{tabular}{cccc}
         ID & Type & Reference & Software\\
         \hline\hline
         VFF1990  & Vashishta & \cite{ff_Vashishta} & LAMMPS\\
         ReaxFF2023  & ReaxFF & \cite{NaoNumJeo2023} & LAMMPS\\
         PoreMS  & Lennard-Jones & \cite{KraRybHol2021, AbrIgl2019, GulTho2006} & GROMACS\\
    \end{tabular}
    \caption{Type of each potential and reference. We will use ID to refer to each potential.}
    \label{tab: forcefields}
\end{table}

The Vashishta implementation in LAMMPS has the following functional form:

\begin{equation}
\begin{array}{lcl}
  V_{\mathrm{Vash}}(\vec{r}_{ij})=\sum_i^{N} \sum_{j>i}^{N} V^{(2)}_{ij}(r_{ij})+\sum_i^{N} \sum_{j\neq i}^{N} \sum_{k > j,k\neq i}^{N} V^{(3)}_{ijk}(r_{ij},r_{ik},\theta_{ijk})\\
  V^{(2)}_{ij}(r_{ij})=\frac{H_{ij}}{r^{\eta_{ij}}}+\frac{Z_iZ_j}{r}\mathrm{e}^{-\frac{r}{\lambda_{1}}}-\frac{D_{ij}}{r^4}\mathrm{e}^{-\frac{r}{\lambda_{4}}}-\frac{W_{ij}}{r^6},\\
  V^{(3)}_{ijk}(r_{ij},r_{ik},\theta_{ijk})=B_{ijk}\frac{[\cos \theta_{ijk}-\cos \theta_{0}]^2}{1+C}[\cos \theta_{ijk}-\cos \theta_{0]^2}\mathrm{e}^{\frac{\gamma_{ij}}{r_{ij}-r_{0}}}\mathrm{e}^{\frac{\gamma_{ik}}{r_{ik}-r_{0,ik}}}, \label{eq: vashishta}\\
   r<r_{c,ij}, \; r_{ij} < r_{0,ij}, \; r_{ik} < r_{0,ik},
\end{array}
\end{equation}
where $\vec{r}_{ij}$ is the separation between atom $i$ and $j$, $r=r_{ij}=\vert \vec{r}_{ij}\vert $ is the distance between the atoms, $N$ is the total number of atoms, $V^{(2)}$ and $V^{(3)}$ are two- and three-body terms respectively. The parameter $H_{ij}$ is the strength of the short-range steric repulsion, $\eta_{ij}$ are the exponents of the steric repulsion term, $Z_i$ is the effective charge, $D_{ij}$ is the strength of the charge–dipole attraction, $W_{ij}$ is the van der Waals interaction strength, and $\lambda_1$ and $\lambda_4$ are the screening lengths for Coulomb and charge–dipole terms, respectively. For the three-body term, $B_{ijk}$ is the strength of the three-body interaction, $\theta_{ijk}$  is the angle between $\vec{r}_{ij}$ and $\vec{r}_{ik}$, $\theta_0$ is the covalent bond angle, and $C$ is the three-body saturation parameter. 

The ReaxFF \cite{ReaxFF2001} is a bond-order-based family of potentials, allowing for continuous bond formation and breaking. The potential has the following functional form: 
\begin{equation}
\begin{array}{lcl}
    V_{\mathrm{reaxff}}(\vec{r}_{ij})&=&V_{bond}(\vec{r}_{ij})+ 
    V_{\mathrm{over}}(\vec{r}_{ij})+
    V_{\mathrm{under}}(\vec{r}_{ij})+\\
    &&V_{\mathrm{val}}(\vec{r}_{ij},\vec{r}_{jk},\Theta_{ijk})+
    V_{\mathrm{pen}}(\vec{r}_{ij},\vec{r}_{jk})+\\
    &&V_{\mathrm{tors}}(\vec{r}_{ij},\vec{r}_{jk},\vec{r}_{kl},\Theta_{ijk},\Theta_{jkl})+\\
    &&V_{\mathrm{conj}}(\vec{r}_{ij},\vec{r}_{jk},\vec{r}_{kl},\Theta_{ijk},\Theta_{jkl})+\\
    &&V_{\mathrm{vdWaals}}(\vec{r}_{ij})+
    V_{\mathrm{Coulomb}}(\vec{r}_{ij}),    \label{eq: reaxff}
\end{array}
\end{equation}
where the right-hand side terms are the bond energy, over-coordination correction, under-coordination correction, valence angle terms, the torsion angles energy, conjugation effects, van der Waals and Coulomb interactions, respectively. The indices $i,j,k,l$ denote four distinctive atoms. The angle between atoms $i,j$ and $k$ is denoted by $\Theta_{ijk}$. The functional forms and parameters are given in \cite{ReaxFF2001}. Regarding functional form and computational overhead, ReaxFF is significantly more complex than Lennard-Jones or Buckingham potentials. It requires hundreds of parameters. The parameters are fitted using DFT calculations and experimental data. Typically, a ReaxFF provides quantum-mechanical accuracy and classical MD scalability.  

The Lennard-Jones (LJ) type combines a potential given by:
\begin{equation}
    V_{\mathrm{LJ}}(\vec{r}_{ij}):=4 \epsilon_{ij} \left [ \left( \frac{\sigma_{ij}}{r_{ij}}\right)^{12} - \left( \frac{\sigma_{ij}}{r_{ij}}\right)^{6} \right], \label{eq: LJ}
\end{equation}
and a Coulomb potential:
\begin{equation}
V_{\mathrm{c}}(\vec{r}_{ij}):=\frac{q_iq_j}{\epsilon r_{ij}}, 
\end{equation} \label{eq: Coulomb}
where $\vec{r}_{ij}$ is the separation between atom $i$ and $j$, $r_{ij}=\vert \vec{r}_{ij}\vert $ is the distance between the atoms, $\epsilon$ is the dielectric constant, $\sigma_{ij}$ and $\epsilon_{ij}$ are parameters, $q_i$ and $q_j$ are the electric charges. All the parameters are included in a Git repository \cite{GalavizGit2025}. 

\subsection{Analysis}

 Regarding the PDOS, each sampling of LAMMPS and GROMACS trajectories consisted of 5000 position snapshots spaced by \qty{0.01}{ps}. For LAMMPS, we use second-order interpolation to generate the velocities. GROMACS does not require a velocity calculation, since the output file already includes it. The total PDOS was weighted using the \texttt{b\_incoherent} and \texttt{b\_coherent} options. The PDOS is normalized by the total degrees of freedom:
\begin{align}
    \mathrm{PDOS}(E)=&3N\frac{\mathrm{pdos}(E)}{||\mathrm{pdos}(E_{\mathrm{max}})||}, \label{eq: dos_normalization}\\
        ||\mathrm{pdos}(E_{\mathrm{max}})||:=&\int_{0}^{E_{\mathrm{max}}}|\mathrm{pdos}(E)| \mathrm{d}E, \nonumber
\end{align}
    
where $\mathrm{pdos}(E)$ is the PDOS as calculated by MDANSE and $N$ is the total number of atoms in the  \ce{SiO2} unit cell. We used \eqref{eq: dos_normalization} when comparing PDOS from MD simulations, as the PDOS is calculated by the same software and under equivalent conditions. 

We also normalize the PDOS by its maximum intensity:
\begin{align}
    \mathrm{PDOS}(E)=&\frac{\mathrm{pdos}(E)}{\lceil{\mathrm{pdos}(E_{\mathrm{max}})\rceil}}. \label{eq: dos_max_normalization}\\
    \lceil{\mathrm{pdos}(E_{\mathrm{max}})}\rceil:=&\max_{E\in [0,E_{\mathrm{max}}]}\left [\mathrm{pdos}(E)\right]. \nonumber
\end{align}
 We used \eqref{eq: dos_max_normalization} when comparing with experimental and DFT calculations, as the PDOS is calculated using different methods, and the resulting PDOS is more convenient for visualization.  

The neutron scattering function was calculated as
\begin{equation}
    S(E) = \frac{PDOS(E)}{E}.
\end{equation}

The axial distribution histogram (ADH) was calculated using a custom C++ program \cite{MDTools} that reads both LAMMPS and GROMACS trajectories. The program calculates the center of mass for each trajectory frame and computes the distance of each atom to a given axial line. The histogram aggregates the estimated distance for each atom at each trajectory frame. The result accounts for every axial location visited by each species during the simulation. 

The structures were visualized using VMD \cite{HumDalSch1996} and VESTA \cite{MomIzu11}.

\subsubsection{Surface vibrations}
\label{app: surface vibrations}

A useful metric for characterizing the vibration of surface atoms is the mode participation ratio \cite{BellDea70, HafKra93, BurKolBru96}:

\begin{equation}
    PR_k=\frac{(\Sigma_i \vert \mathbf{e}_{i,k} \vert^2)^2}{N\Sigma_i \vert \mathbf{e}_{i,k} \vert^4},
\end{equation}
where $\mathbf{e}_{i,k}$ is the normalized eigenvector component for atom $i$, and $N$ is the total number of atoms. The required eigenvector components are usually calculated from the lattice dynamics. Since our simulations employ molecular dynamics, we calculated the phonon participation ratio (PPR) instead \cite{MeyCom11, LohTeoTay12, LiaZhoZha20, ZhoLiaWu20, XuLiaZha22}. Using a per-atom vibrational density of state $VDOS_i(E)$, we can define a normalized distribution of modes:
\begin{equation}
    p_i=\frac{VDOS_i(E)}{\sum_i^N VDOS_i(E)}.
\end{equation}
We can then define an Inverse Participation Ratio (IPR):
\begin{equation}
    IPR=\sum_i^N p_i^2.
\end{equation}
In quantum mechanics, the IPR is defined as the integral (or sum for a discrete system) over the square of the wave function density. In this case, values close to 1 correspond to localized states (a few atoms contribute strongly to the vibrational density at a given energy), and values close to 1/N correspond to delocalized states (every atom contributes equally to the vibrational density). Therefore, we can renormalize the IPR by defining PPR as:  
\begin{equation}
PPR(E)=\frac{1}{N\sum_i^N p_i^2}.
\end{equation}
Notice that with this definition, for $N \gg 1$, values close to $1/N\rightarrow0$ correspond to localized states, and values close to 1 are delocalized ones.

In addition to the PPR, we calculate an \ce{H2O} fraction contribution to the VDOS:
\begin{equation}
    f_{\ce{H2O}}(E)=\frac{\sum_{i\in \ce{H2O}} VDOS(E)_i}{\sum_i^N VDOS(E)_i},
\end{equation}
where the numerator is the sum of all the per-atom vibration modes from the \ce{H2O} atoms, and the denominator includes all the atoms. In this case, $f_{\ce{H2O}}(E) \rightarrow 1$ when the majority of the VDOS weight at a given energy originates from the \ce{H2O} atoms. Conversely, $f_{\ce{H2O}}(E) \rightarrow 0$ when the \ce{H2O} contribution is low.

\subsection{Density functional theory calculation}

Initial atomic positions from the $\alpha-$quartz structure were used to generate input files for QE. The files were generated using the Materials Cloud platform \cite{TalKumPas2020}. We utilized the PBEsol generalized gradient approximation for the exchange-correlation functional \cite{PerRuz2008}. The electron-ion interactions were modeled using PAW pseudopotentials from the SSSP library \cite{QE_Pot1,QE_Pot2}. We performed an energy cutoff test (\texttt{ecutwfc} parameter in QE) with energies between \qty{60}{Ry} and \qty{110}{Ry}. We selected \ce{75}{Ry} as the optimal value, with a relative change in the calculated energy of less than 0.001\% at 110 Ry. Similarly, we performed a k-point convergence test, choosing a $10\times 10 \times 8$ mesh without offset. We used those parameters for relaxation and phonon calculations. We performed variable cell relaxation using damped (quick-min Verlet) dynamics, a 0.00001 force convergence threshold (\texttt{forc\_conv\_thr}), and a 0.000009 energy convergence threshold (\texttt{etot\_conv\_thr}). For phonon calculation, we generated the displacement structures using \textsc{Phonopy} on a $2\times 2 \times 2$ supercell, resulting in nine perturbed structures. The PDOS was calculated from \qty{0}{meV} to \qty{150}{meV} with a \qty{0.5}{meV} resolution and \qty{5}{meV} Gaussian broadening width.

\section{Results}
\label{sec: results}

\subsection{Force field dependency}
\label{sec: force_field_dependency}

\begin{figure}[!ht]
\centering
    \includegraphics[width=\textwidth]{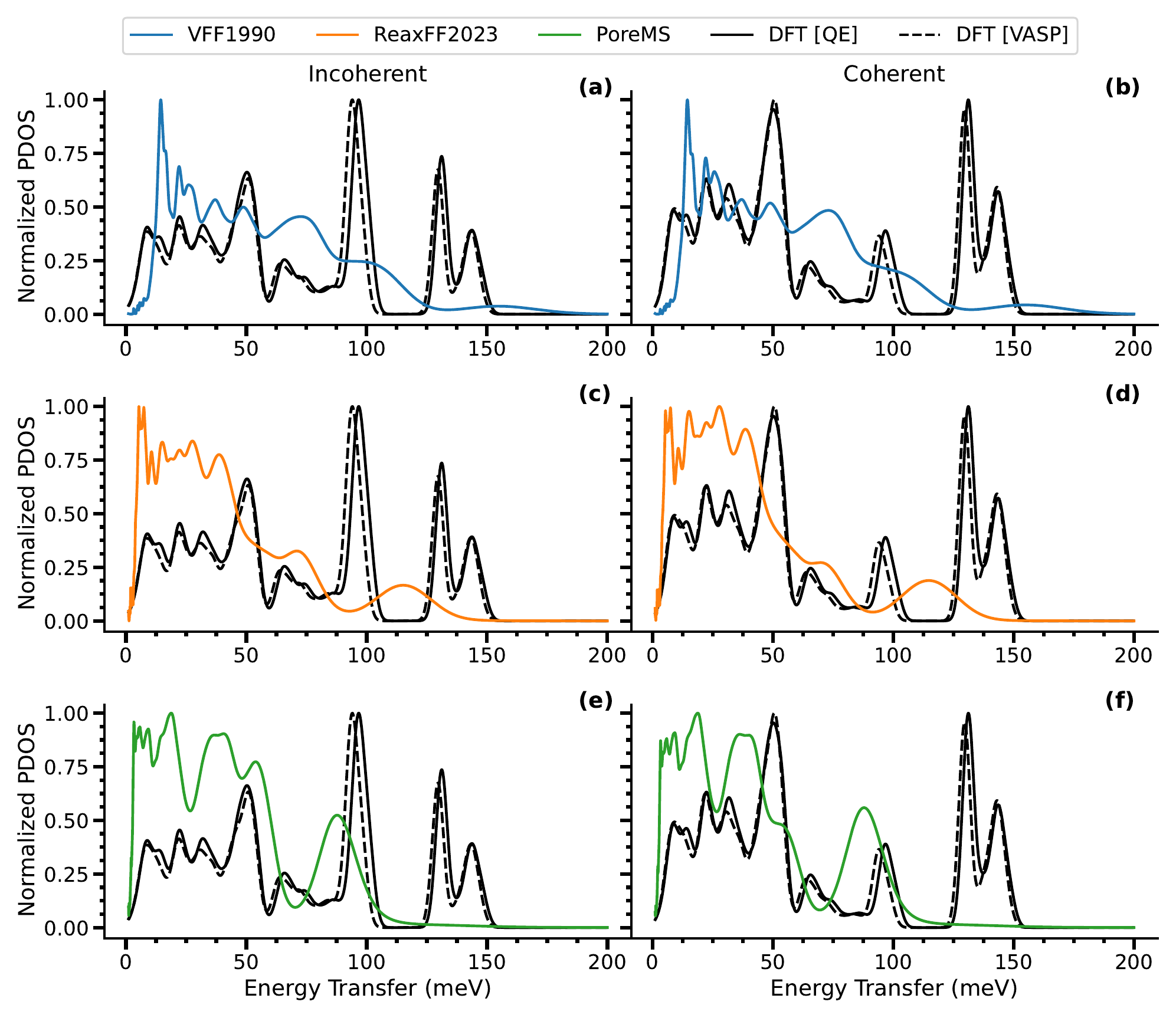}
\caption{Phonon density of states for $\alpha$-quartz comparison using three force fields and DFT calculations. Panels \textbf{(a)}-\textbf{(b)} show the comparison between VFF1990 and the DFT calculations for Incoherent and Coherent scattering, respectively. Panels \textbf{(c)}-\textbf{(d)}, and \textbf{(e)}-\textbf{(f)} shows the corresponding result for ReaxFF2023 and PoreMS force fields, respectively.} \label{fig: alpa_quartz_dft_comparison}
\end{figure}

We calculated the PDOS for $\alpha$-quartz using three force fields and compared the results with DFT calculations performed using the QE and VASP software. The MD result is from a \qty{300}{K} simulation. The DFT results are consistent with those from other calculations in the literature \cite{MurPre2021}. Figure \ref{fig: alpa_quartz_dft_comparison} shows the result. For this structure, the three force fields fail to reproduce the PDOS peaks between \qty{100}{meV} and \qty{150}{meV}. The PoreMS force field appears to be most consistent with the DFT calculations in the lower-energy range. 

\begin{figure}[!ht]
\centering
    \includegraphics[width=\textwidth]{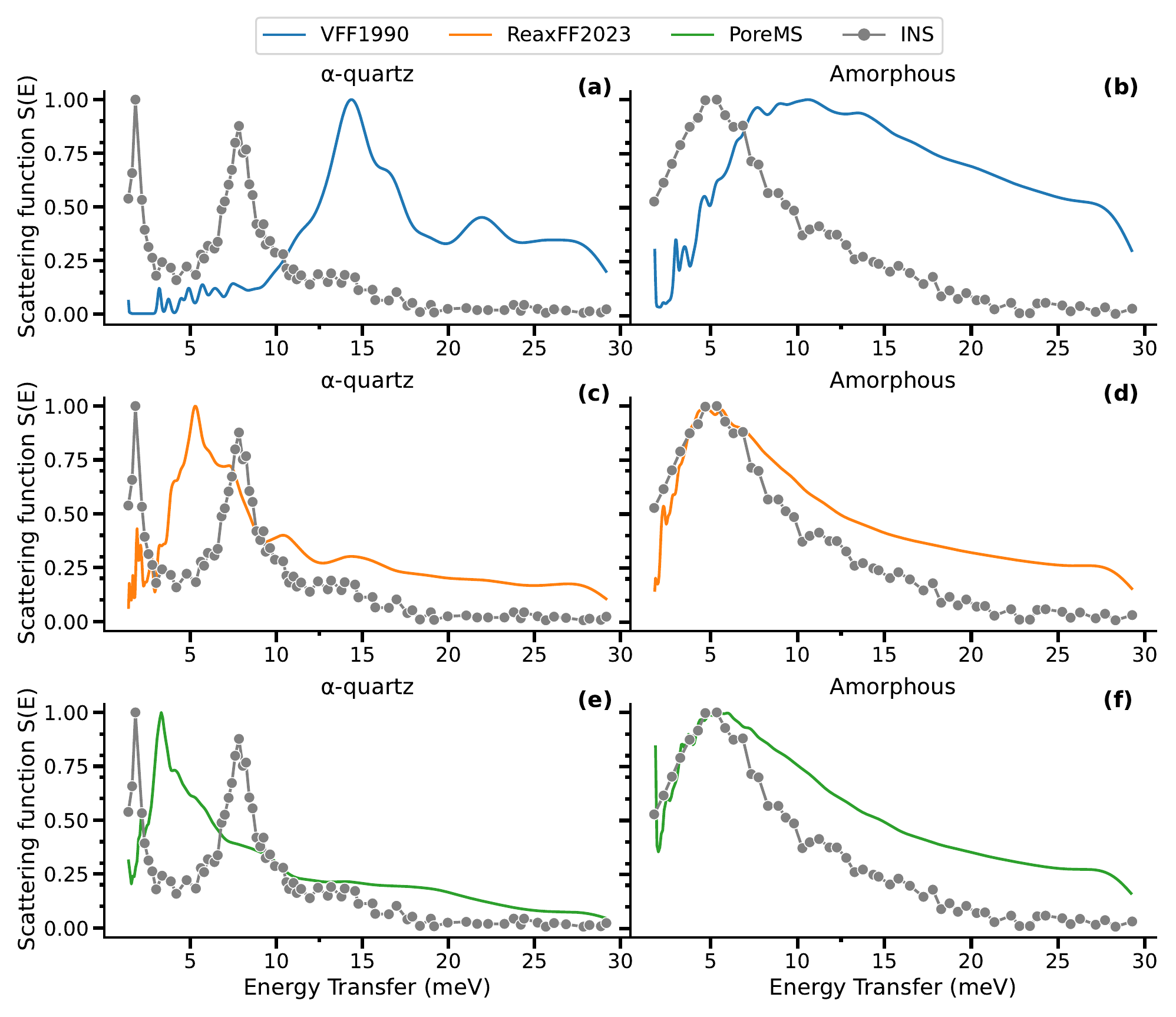}
\caption{Neutron scattering function for $\alpha$-quartz and amorphous \ce{SiO2} comparison using three force fields and INS results. Panels \textbf{(a)}-\textbf{(b)} show the comparison between the INS experiment and the calculated VFF1990 for $\alpha$-quartz and amorphous \ce{SiO2}, respectively. Panels \textbf{(c)}-\textbf{(d)}, and \textbf{(e)}-\textbf{(f)} shows the corresponding result for ReaxFF2023 and PoreMS force fields, respectively.} \label{fig: INS_comparison}
\end{figure}

We performed a comparison between the calculated $\alpha$-quartz and amorphous \ce{SiO2} neutron scattering function and the respective INS results from \cite{HarDovPar2000}.
The inelastic neutron scattering data were obtained using the PRISMA time-of-flight neutron spectrometer at the ISIS Facility (see reference for details). The data was extracted from the published figures using an online tool \cite{PlotDigitalizer}. Figure \ref{fig: INS_comparison} shows the results. Note that the experimental data only show the low-energy range (up to \qty{30}{meV}). The simulation and the experiment are for a system at \qty{300}{K}. 

In the case of $\alpha$-quartz, the experimental data show one peak at \qty{1.8}{meV} and a second one at \qty{7.8}{meV}. In contrast the VFF1990 shows a main peak at around \qty{15}{meV} and a secondary small peak at \qty{22.5}{meV} (see \ref{fig: INS_comparison}-\textbf{(a)}). Therefore, it seems that there is a shift to higher energy. The ReaxFF2023 and PorMS force fields show the opposite results. There is a single peak at around \qty{5}{meV} and \qty{3.5}{meV}, respectively (see \ref{fig: INS_comparison}-\textbf{(c)} and \textbf{(e)}).

In the case of amorphous \ce{SiO2}, the experiment shows a single peak at around \qty{5}{meV} which is well modeled by using the ReaxFF2023 and PoreMS force fields (see \ref{fig: INS_comparison}-\textbf{(d)} and \textbf{(f)}). The VFF1990 once again shows an energy shift to larger values with a broader peak at around \qty{10}{meV} (see \ref{fig: INS_comparison}-\textbf{(b)}).

\subsubsection{Temperature dependency}
\label{sec: FF_temperature_dependency}

\begin{figure}[!ht]
\centering
    \includegraphics[width=\textwidth]{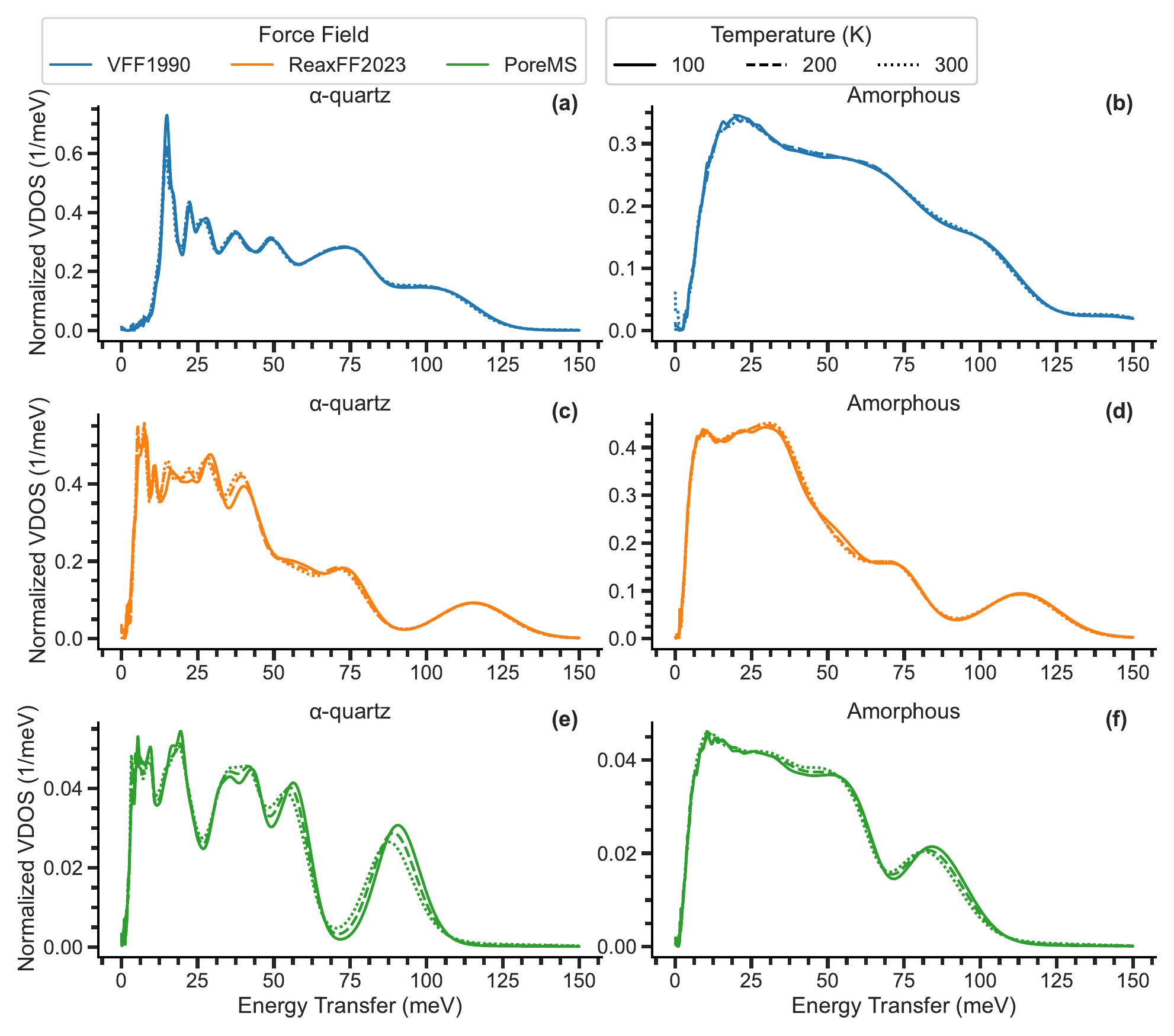}
\caption{VDOS temperature dependency for $\alpha$-quartz and amorphous \ce{SiO2}. Panels \textbf{(a)}-\textbf{(b)} show the VDOS temperature dependency using VFF1990 for $\alpha$-quartz and amorphous \ce{SiO2}, respectively. Panels \textbf{(c)}-\textbf{(d)}, and \textbf{(e)}-\textbf{(f)} show the corresponding result for ReaxFF2023 and PoreMS force fields, respectively. In each panel, the temperature is indicated by the line style.} \label{fig: FF_temperature_dependency}
\end{figure}

We tested the VDOS dependence on temperature at \qty{100}{K}, \qty{200}{K}, and \qty{300}{K}. The results are presented in Figure \ref{fig: FF_temperature_dependency}. For VFF1990, temperature has a minor effect over the tested temperature range. The ReaxFF20213 result shows VDOS broadening at energies below \qty{50}{meV} for the $\alpha$-quartz simulation and only a minor effect for the amorphous ones. The PoreMS result shows the most pronounced broadening effect across the entire energy range and for both structures. Overall, the VDOS is primarily in a harmonic regime, with only minor structural changes observed over the simulated energy range. 

\subsection{Nanopore vs bulk comparison}
\label{sec: Nanopore_vs_bulk_comparison}

\begin{figure}[!ht]
\centering
    \includegraphics[width=\textwidth]{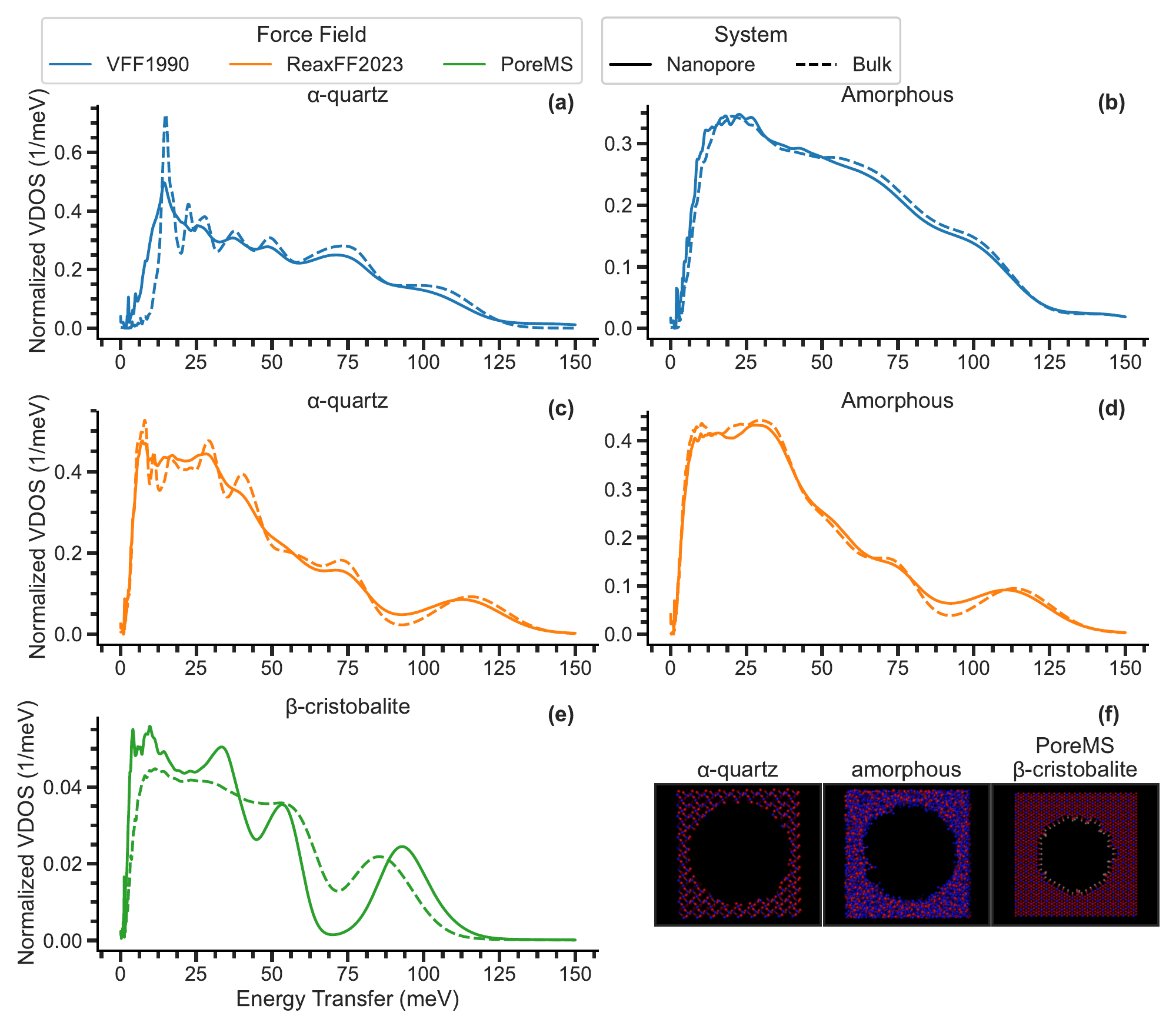}
\caption{VDOS for $\alpha$-quartz and amorphous \ce{SiO2} nanopore at \qty{300}{K}. Panels \textbf{(a)}-\textbf{(b)} show the VDOS of a $r=\qty{2}{nm}$ nanopore using VFF1990 for $\alpha$-quartz and amorphous \ce{SiO2}, respectively. Panels \textbf{(c)} and \textbf{(d)} show the corresponding result for ReaxFF2023. Panel \textbf{(e)} shows the result for $\alpha$-quartz using the PoreMS force field. In each panel, the nanopore is represented by the solid line, and the dashed line represents the bulk structure simulation. Panel \textbf{(f)} shows the initial structure; the left and middle structures are used for LAMMPS simulations, and the right one for GROMACS. } \label{fig: FF_pore}
\end{figure}

We performed simulations of a $r=\qty{2}{nm}$ nanopore for temperatures \qty{100}{K}, \qty{200}{K}, and \qty{300}{K}. For simulations using VFF1990 and ReaxFF2023, the pore consisted of a bulk structure where atoms inside a cylinder of radius \qty{2}{nm} were removed (see Figure \ref{fig: FF_pore}-\textbf{(f)}). The standard VFF1990 does not include \ce{OH} binding.\footnote {We attempted to use a hybrid force field with LJ parameters for oxygen and hydrogen binding.} Therefore, the nanopore does not include silanol groups. Similarly, when using ReaxFF2023, we encounter issues when the simulation includes silanol groups. We were unable to stabilize the system, and atoms were lost during the NVT equilibration phase. In contrast, the structure generated with PoreMS is stable. The issue persisted when using the PoreMS structure in LAMMPS. Therefore, in the following, the results for the LAMMPS simulation don't include silanol groups. Similarly, we encounter issues when simulating an amorphous pore with GROMACS. Therefore, all the PoreMS results are for the $\alpha$-quartz nanopore. 

The inclusion of the cavity has a significant effect on the $\alpha$-quartz VDOS (see Figure \ref{fig: FF_pore} panels \textbf{(a)}, \textbf{(c)} and \textbf{(e)}). In the VFF1990 simulation, the main change occurs at lower-energy modes. On one hand, the VDOS at \qty{0}{meV} increases more rapidly, showing additional peaks below \qty{5}{meV}. The peaks are broadened, in particular the one near \qty{12.5}{meV}. In the ReaxFF2023 simulation, the peaks are similarly broadened, with some low-energy modes merging. The most significant change occurs for the PoreMS simulation. There is a combined effect of peak broadening and shifting, with most modes shifting to lower energies. 

For the amorphous pore VDOS, the cavity effect is less significant. For both force fields (VFF1990 and ReaxFF2023), we observe a slight broadening, and for VFF1990, some additional peaks appear at very low energies.

\subsection{\ce{H2O} loading dependency}
\label{sec: H2O_content_dependency}

We investigated the effect of adding \ce{H2O} molecules inside the nanopore using ReaxFF2023 and PoreMS with an implementation of a TIP4P2005 flexible water model for GROMACS\footnote{In LAMMPS, we tried with a hybrid force field using VFF1990 and a TIP4P2005 water model. However, the resulting simulation crashes during the heating phase. }. The \ce{H2O} molecules are initially randomly located inside the nanopore. After the initial heating phase, the \ce{H2O} molecules uniformly distribute on the pore's surface (see \ref{app: heating_procedure_test}). 

\begin{figure}[!ht]
\centering
    \includegraphics[width=0.9\textwidth]{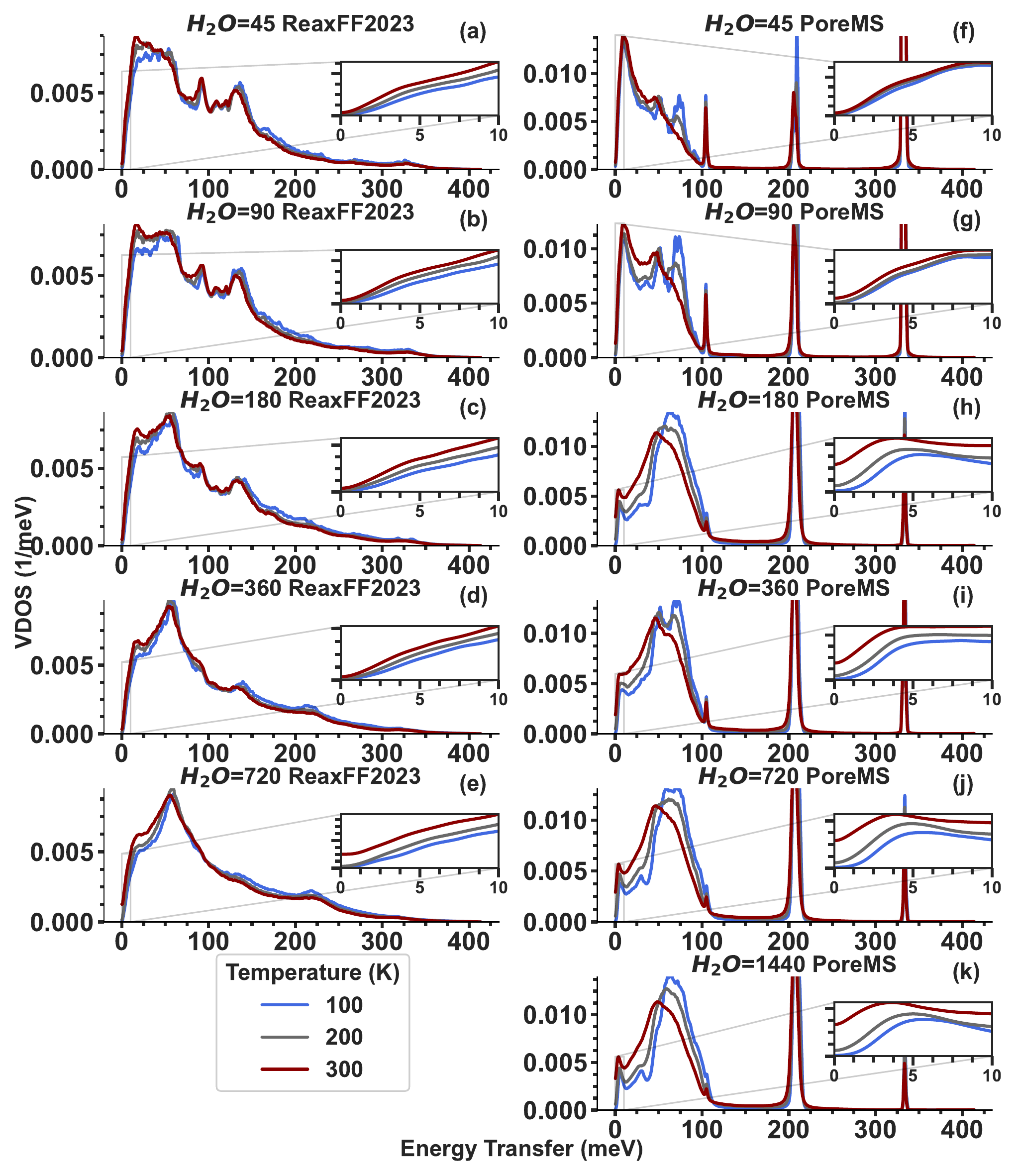}
\caption{VDOS for an $r=\qty{2}{nm}$ nanopore with \ce{H2O} molecules. Panels \textbf{(a)}-\textbf{(e)} show the VDOS for a simulation using ReaxFF2023 with 45, 90, 180, 360 and 720 \ce{H2O} molecules. Panels \textbf{(f)}-\textbf{(k)} show the corresponding result for the PoreMS force field with an additional simulation with 1440 \ce{H2O} molecules.} \label{fig: added H2O}
\end{figure}

Figure \ref{fig: added H2O} shows the VDOS for a r=\qty{2}{nm} nanopore. We calculated the VDOS at three temperatures \qty{100}{K}, \qty{200}{K} and \qty{300}{K}. We simulated systems with 45, 90, 180, 360, 720, and 1440 \ce{H2O} molecules. The left panels show the results obtained with the ReaxFF2023 force field. The simulation with 1440 \ce{H2O} molecules is missing because it was unstable. The main effect of adding \ce{H2O} molecules is the appearance of a dominant mode at around \qty{60}{meV}. The effect of temperature is more pronounced, showing a broader VDOS at higher temperatures. This is particularly clear for energies below \qty{50}{meV}. The intercept of the VDOS at \qty{0}{meV} shows a noticeable increase of modes (see Figure \ref{fig: added H2O} insets). The right panels show the results obtained with the PoreMS force field using the TIP4P2005f water model. The overall result is similar; adding water shifts the low-energy modes to a broad peak near \qty{60}{meV}. However, in this case, the position of the dominant peak depends on temperature. In addition, we observe the development of secondary peaks near \qty{200}{meV} and \qty{330}{meV}. In Section \ref{sec: H2O_monolayer_analysis}, we will show that the first peak originates from the TIP4P2005 water model and the second from the nanopore. However, here we see that the intensity of the \qty{200}{meV} peak increases with \ce{H2O} load, while the peak at \qty{330}{K} decreases. 

\begin{figure}[!ht]
\centering
    \includegraphics[width=\textwidth]{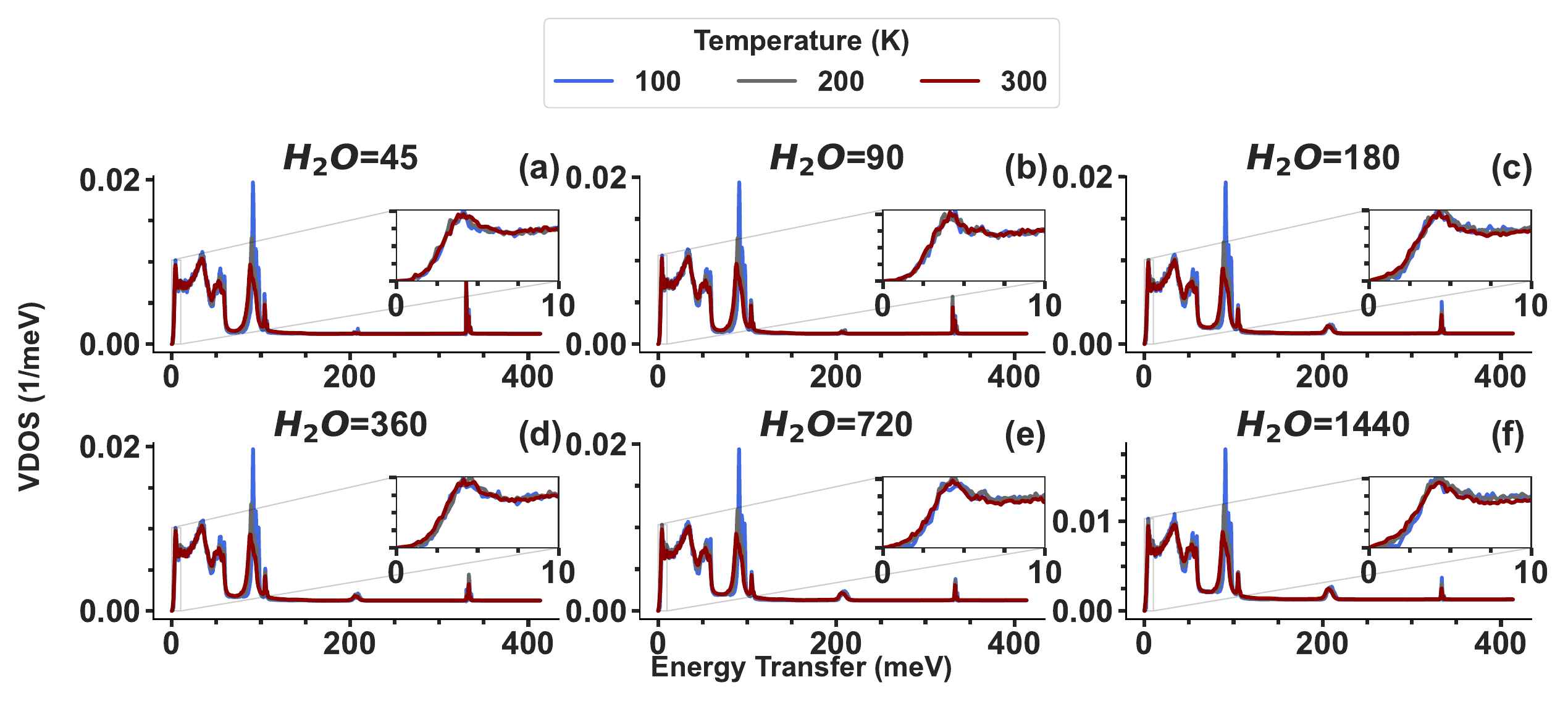}
\caption{GROMACS calculated VDOS for an $r=\qty{2}{nm}$ nanopore with \ce{H2O} molecules. Panels \textbf{(a)}-\textbf{(f)} show the VDOS for a simulation using the PoreMS force field with 45, 90, 180, 360, 720, and 1440 \ce{H2O} molecules. The insets show the corresponding low-energy range for the same simulations.} \label{fig: gmx dos}
\end{figure}

For PoreMS, we calculated the VDOS using the GROMACS command \texttt{gmx dos}. Figure \ref{fig: gmx dos} shows the VDOS and the root mean square deviation (RMSD) calculated with the \texttt{gmx rms} command. In GROMACS, the phonon density of states is weighted by mass, while the calculation in MDANSE is weighted by incoherent scattering. Therefore, the changes in the density of states with the addition of \ce{H2O} molecules are less pronounced. We observe a peak near \qty{200}{meV}, but at much lower intensity. Additionally, the low-energy range exhibits a smooth curve near \qty{0}{meV} with a temperature-dependent slope more pronounced at higher water loads (see Figure \ref{fig: gmx dos} \textbf{(g)}-\textbf{(l)}). 

\begin{figure}[!ht]
\centering
    \includegraphics[width=\textwidth]{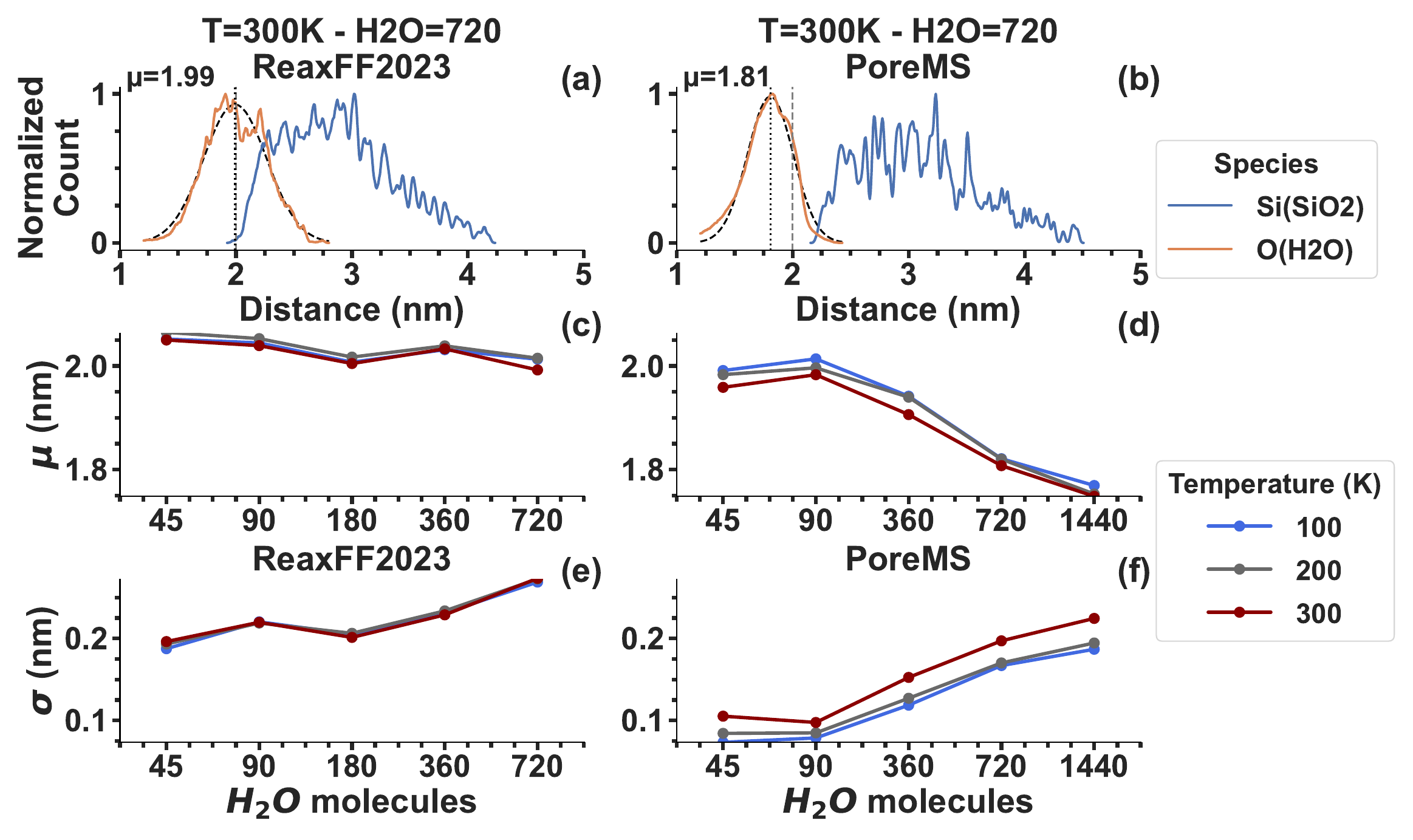}
\caption{Axial distribution for an $r=\qty{2}{nm}$ nanopore with \ce{H2O} molecules. Panels \textbf{(a)} and \textbf{(b)} show the ADH at \qty{300}{K} with 720 \ce{H2O} molecules loading for the ReaxFF2023 and PoreMS simulations, respectively. Panels \textbf{(c)} and \textbf{(d)} show the mean of a Gaussian distribution fitted to the ADH of the \ce{H2O} oxygens as a function of temperature and \ce{H2O} loading for the ReaxFF2023 and PoreMS simulations, respectively. Panels \textbf{(e)} and \textbf{(f)} show the standard deviation of a Gaussian distribution fitted to the ADH of the \ce{H2O} oxygens as a function of temperature and \ce{H2O} loading for the ReaxFF2023 and PoreMS simulations, respectively.} \label{fig: H2O sigma mu}
\end{figure}

Figure \ref{fig: H2O sigma mu} shows the axial distribution. The top panels illustrate the calculation. The plot shows the distribution of oxygen from the \ce{H2O} molecule and the silicon from the pore relative to the center of the pore. The vertical line at \qty{2}{nm} indicates the expected pore radius. The silicon shows a flat profile from around \qty{2}{nm} to \qty{3}{nm} and then a lineal decay from \qty{3}{nm} to $3\sqrt{2}$nm. This is expected from the pore's geometry.
On the other hand, the \ce{H2O} molecules are distributed on the pore's surface. We fitted a Gaussian distribution (indicated with a dashed line). The corresponding mean is shown in the figure by a vertical dotted line, and its value is given in the top-left corner of the plots. Note that the ReaxFF2023 and PoreMS simulations exhibit distinct behavior. The reactive nature of ReaxFF2023 is reflected in the more pronounced overlap between the oxygen and silicon atoms. This suggests that there is some interchange of oxygen atoms between the \ce{H2O} and the \ce{SiO2}. In contrast, the PoreMS simulation shows a more precise separation between the \ce{H2O} and the \ce{SiO2}, with the mean of the \ce{H2O} distribution closer to the center of the pore.   

Figure \ref{fig: H2O sigma mu} panels \textbf{(c)} to \textbf{(f)} show the fitted mean and standard deviation as a function of \ce{H2O} molecule loading and temperature. The ReaxFF2023 results (left panels) show a weak temperature dependence and a significant interaction with the pore, as evidenced by the nearly constant mean and a comparatively large standard deviation that increases with \ce{H2O} loading. The PoreMS result shows that the water molecules form a thicker layer on the pore wall with the mean of the distribution shifting away from the pore's wall and the standard distribution increasing with increasing \ce{H2O} load. 

\subsection{\ce{H2O} monolayer analysis}
\label{sec: H2O_monolayer_analysis}

For the $r=\qty{2}{nm}$ nanopore, a loading of 720 \ce{H2O} molecules forms a monolayer that covers the pore's wall. We analyzed this system in more detail. We performed a simulation over the temperature range \qty{100}{K} to \qty{400}{K} in \qty{50}{K} increments. The simulation uses the protocol outlined in Protocol \ref{protocol: long}.  

\begin{figure}[!ht]
\centering
    \includegraphics[width=\textwidth]{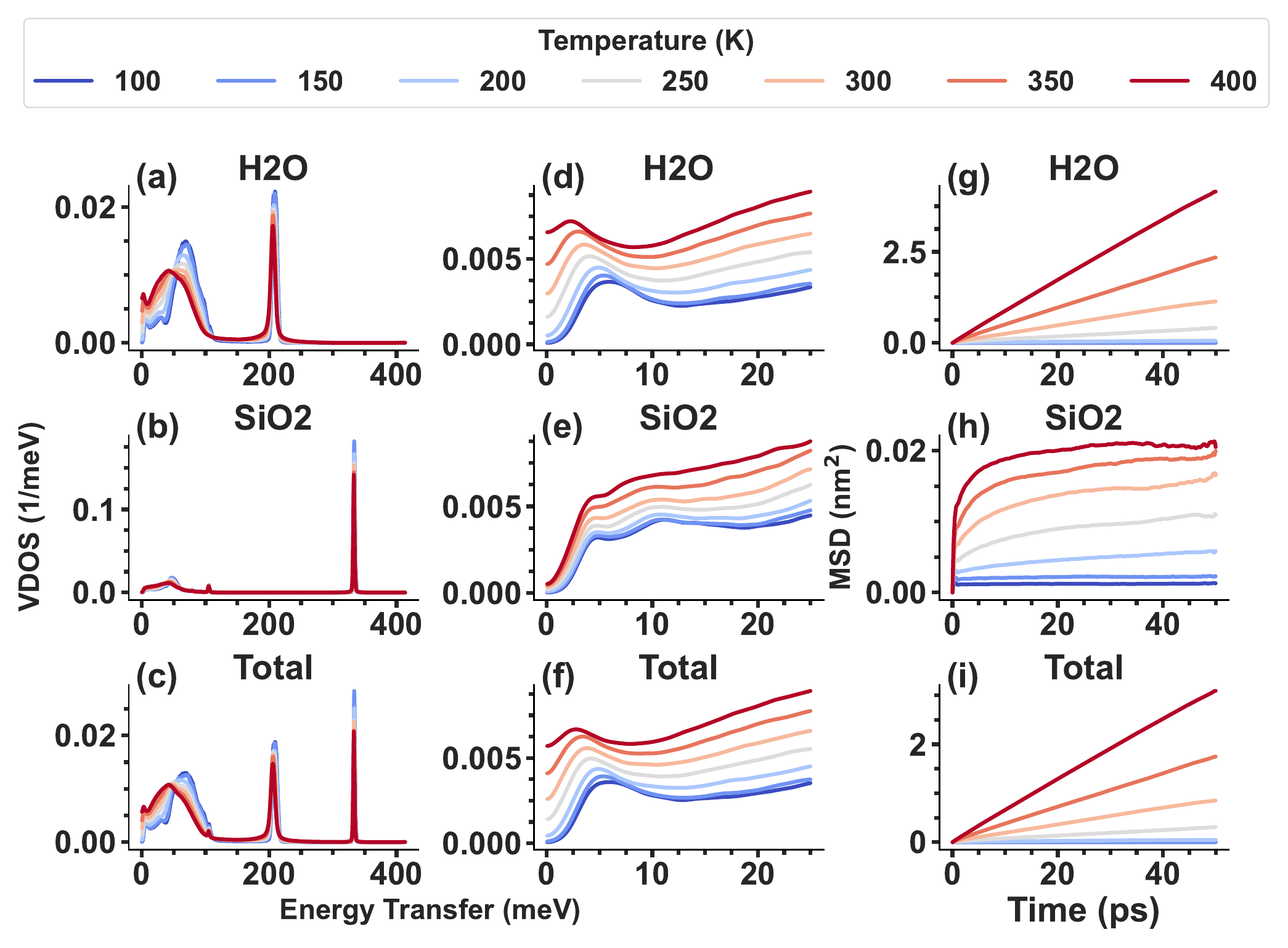}
\caption{VDOS and MSD for an $r=\qty{2}{nm}$ nanopore with 720 \ce{H2O} molecules. Panels \textbf{(a)} and \textbf{(b)} show the partial VDOS for \ce{H2O} and \ce{SiO2} molecules. Panel \textbf{(c)} shows the corresponding total incoherent VDOS. Panels \textbf{(d)} to \textbf{(f)} show the corresponding low-energy range VDOS. Panels \textbf{(g)} and \textbf{(h)} show the MSD of \ce{H2O} and \ce{SiO2} molecules. Panel \textbf{(i)} shows the MSD of the whole system.} \label{fig: 720 H2O dos MSD}
\end{figure}

Figure \ref{fig: 720 H2O dos MSD} shows the VDOS and the MSD. Panels \textbf{(a)} to \textbf{(c)} confirms that the peaks at \qty{100}{meV} and \qty{330}{meV} are due to the \ce{SiO2}. The peak near \qty{200}{meV} comes from the \ce{H2O} vibrations. The intercept at \qty{0}{meV} shows a change in the slope and an intensity increase due to the temperature change (see panels \textbf{(d)} to \textbf{(f)}). The transition between the liquid-like linear scaling and the quadratic Debye law as a function of temperature was previously reported in \cite{JinFanSta2024}. In addition, panels \textbf{(g)} and \textbf{(i)} show an uneven increase in the MSD, indicating diffusion of \ce{H2O} molecules at temperatures higher than \qty{250}{K}. In contrast, panel \textbf{(h)} shows negligible MSD with a more even increase as a function of temperature. Table \ref{tab: Diff Coeff} shows the diffusion coefficients calculated from the slope of a fitted linear function to the MSD\footnote{The fitting is performed for times after \qty{5}{ps}. In that way, we avoid the initial quadratic portion of the \ce{SiO2} MSD}. The diffusion coefficients for temperatures above \qty{200}{K} are comparable with the experimental results.  

\begin{table}[!ht]
\centering
\caption{Diffusion coefficients calculated from MSD. Experimental data from \cite{DiffH2O_WEB,DiffH2O_1,DiffH2O_2,DiffH2O_3,DiffH2O_4,DiffH2O_5,DiffH2O_6,DiffH2O_7,DiffH2O_8,DiffH2O_9}} \label{tab: Diff Coeff}
\begin{tabularx}{0.85\textwidth}{rcccc}
\toprule
 & \multicolumn{4}{c}{D ($\mu m^2/ms$)} \\
Temperature (K) & \ce{H2O} & \ce{SiO2} & Total & Experimental \\
\midrule
100 & 0.00 & 0.00 & 0.00 & 0.00 \\
150 & 0.01 & 0.00 & 0.01 & 0.00 \\
200 & 0.16 & 0.01 & 0.12 & 0.00 \\
250 & 1.25 & 0.01 & 0.94 & 0.34 \\
300 & 3.80 & 0.02 & 2.83 & 2.41 \\
350 & 7.68 & 0.02 & 5.72 & 6.29 \\
400 & 13.81 & 0.01 & 10.28 & 11.26 \\
\bottomrule
\end{tabularx}
\end{table}

\begin{figure}[!ht]
\centering
    \includegraphics[width=\textwidth]{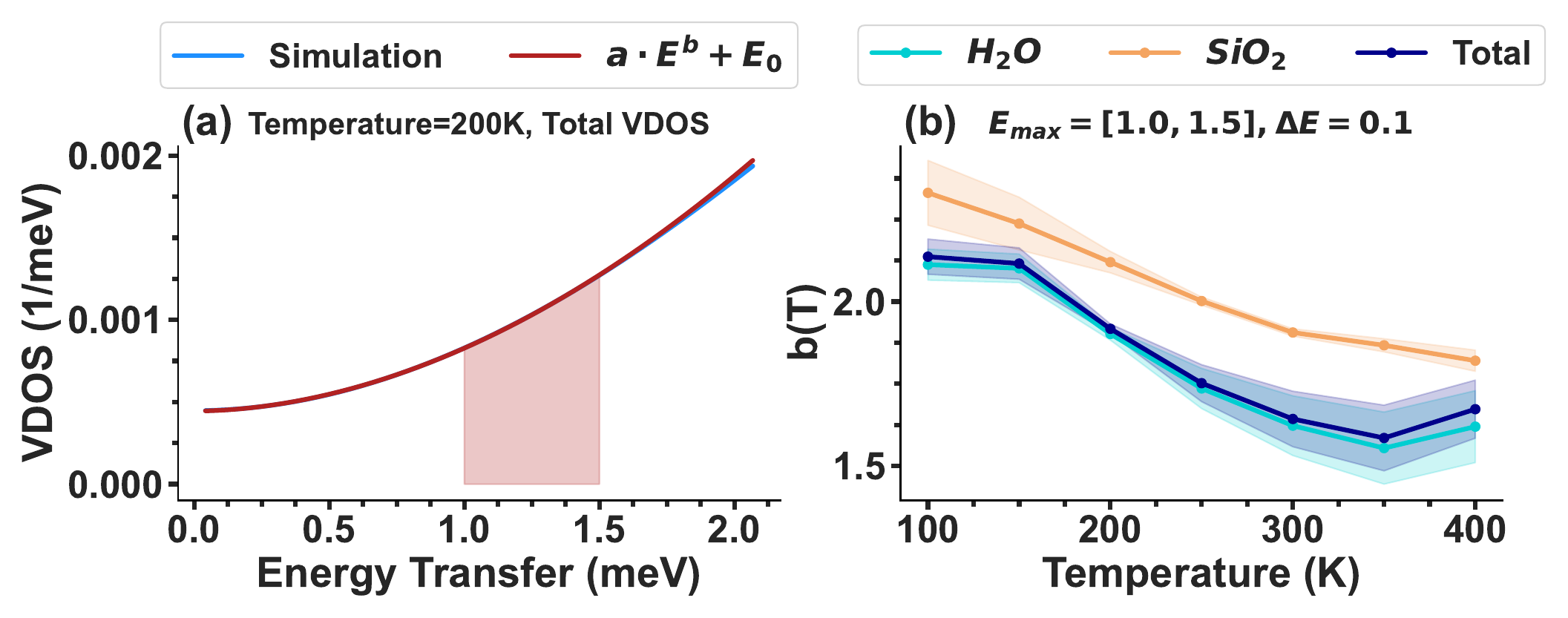}
\caption{Functional form of the low energy VDOS for an $r=\qty{2}{nm}$ nanopore with 720 \ce{H2O} molecules. Panel \textbf{(a)} the energy range is varied, fitting function for the total VDOS and \qty{100}{K}. The shaded area indicates the interval considered for the sensitivity analysis of the fit. Panel \textbf{(b)} shows the mean of the fitted power coefficient $b(T)$ as a function of temperature. The shaded region represents the one-standard-deviation interval for the fit as the energy range varies.} \label{fig: VDOS power}
\end{figure}

We fitted a polynomial $P(E)=a \cdot E^b+E_0$ to the low-energy VDOS. Since the fitting depends on the energy range considered, we performed a sensitivity analysis. We fitted the function to the VDOS with energy between $[0,E_{max}]$ changing $E_{max}$ from \qty{1.0}{meV} to \qty{1.5}{meV} in \qty{0.1}{meV} steps. Figure \ref{fig: VDOS power}-\textbf{(a)} shows a typical fitting where the shading area denotes the maximum energy interval. Figure \ref{fig: VDOS power}-\textbf{(b)} shows the values of the fitted power coefficient $b(T)$. We observe that the low modes transition from a quadratic power law to an almost linear increase in the considered temperature range.  

Table \ref{tab: VDOS fit stats} summarizes the fitting results. The incoherent total VDOS is dominated by the \ce{H2O}. At low temperature, a quadratic scaling predicted by the Debye model is observed, followed by a transition to linear scaling characteristic of liquid diffusion \cite{YuYanBag22}. 

\begin{table}[!ht]
    \centering
    \caption{Mean ($\mu$) and standard deviation ($\sigma$) of the fitting coefficients for each temperature.} \label{tab: VDOS fit stats}
\begin{tabularx}{\textwidth}{rrrcccrc}
\toprule
 &  & \multicolumn{2}{c}{a ($\times 10^4$)} & \multicolumn{2}{c}{b } & \multicolumn{2}{c}{$E_0\;(\times 10^4)$} \\
 &  & $\mu$ & $\sigma$ & $\mu$ & $\sigma$ & $\mu$ & $\sigma$ \\
Temperature (K) & Subset &  &  &  &  &  &  \\
\midrule
\multirow[t]{3}{*}{100} & \ce{H2O} & 1.5 & 0.01 & 2.1 & 0.05 & 0.7 & 0.01 \\
 & \ce{SiO2} & 1.0 & 0.01 & 2.3 & 0.10 & 0.4 & 0.01 \\
 & Total & 1.4 & 0.01 & 2.1 & 0.05 & 0.7 & 0.01 \\
\cline{1-8}
\multirow[t]{3}{*}{150} & \ce{H2O} & 2.0 & 0.01 & 2.1 & 0.04 & 1.3 & 0.01 \\
 & \ce{SiO2} & 1.3 & 0.01 & 2.2 & 0.08 & 0.6 & 0.01 \\
 & Total & 1.9 & 0.01 & 2.1 & 0.05 & 1.2 & 0.01 \\
\cline{1-8}
\multirow[t]{3}{*}{200} & \ce{H2O} & 4.2 & 0.01 & 1.9 & 0.02 & 5.0 & 0.01 \\
 & \ce{SiO2} & 1.8 & 0.00 & 2.1 & 0.03 & 1.1 & 0.01 \\
 & Total & 3.8 & 0.00 & 1.9 & 0.01 & 4.5 & 0.01 \\
\cline{1-8}
\multirow[t]{3}{*}{250} & \ce{H2O} & 7.0 & 0.03 & 1.7 & 0.06 & 15.9 & 0.06 \\
 & \ce{SiO2} & 2.6 & 0.00 & 2.0 & 0.01 & 2.3 & 0.00 \\
 & Total & 6.4 & 0.02 & 1.8 & 0.06 & 14.0 & 0.05 \\
\cline{1-8}
\multirow[t]{3}{*}{300} & \ce{H2O} & 8.2 & 0.05 & 1.6 & 0.09 & 29.5 & 0.11 \\
 & \ce{SiO2} & 3.7 & 0.00 & 1.9 & 0.01 & 3.5 & 0.00 \\
 & Total & 7.5 & 0.04 & 1.6 & 0.08 & 25.8 & 0.09 \\
\cline{1-8}
\multirow[t]{3}{*}{350} & \ce{H2O} & 7.2 & 0.06 & 1.6 & 0.11 & 46.8 & 0.12 \\
 & \ce{SiO2} & 4.7 & 0.01 & 1.9 & 0.02 & 4.6 & 0.01 \\
 & Total & 6.8 & 0.05 & 1.6 & 0.10 & 40.8 & 0.10 \\
\cline{1-8}
\multirow[t]{3}{*}{400} & \ce{H2O} & 2.8 & 0.02 & 1.6 & 0.11 & 65.6 & 0.05 \\
 & \ce{SiO2} & 5.9 & 0.01 & 1.8 & 0.03 & 5.3 & 0.02 \\
 & Total & 3.2 & 0.02 & 1.7 & 0.09 & 57.0 & 0.04 \\
\bottomrule
\end{tabularx}
\end{table}

\begin{figure}[!ht]
\centering
    \includegraphics[width=\textwidth]{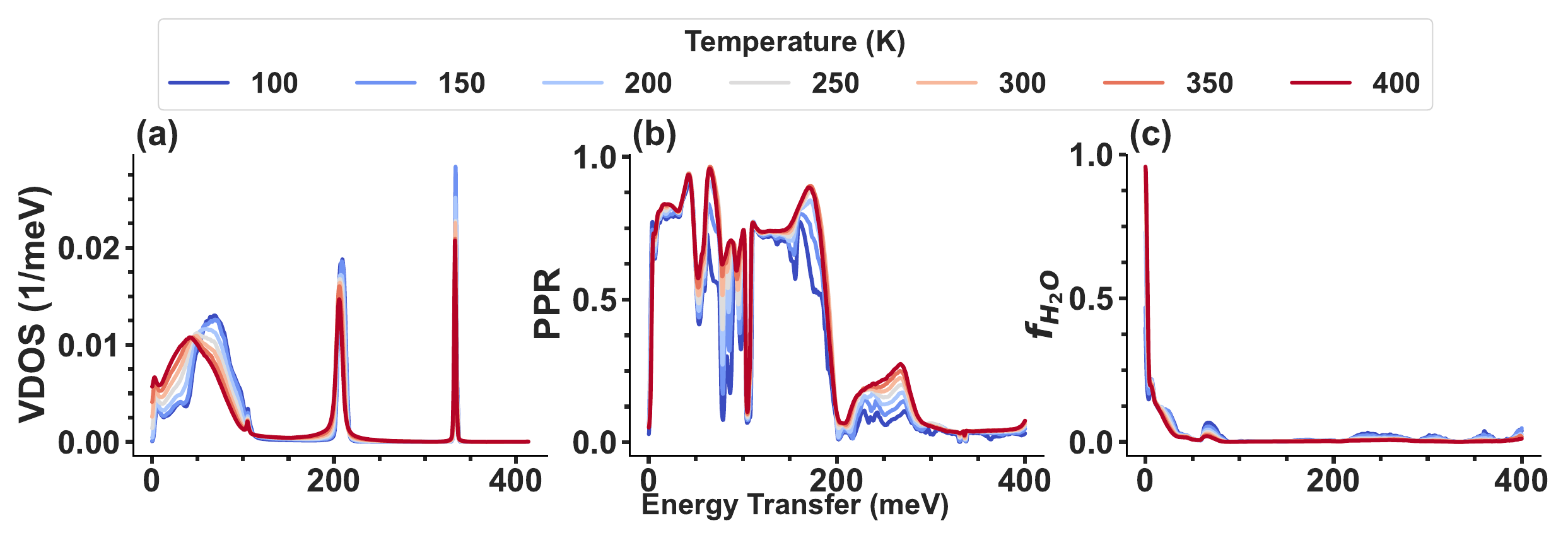}
\caption{Vibrational DOS, phonon participation ratio and \ce{H2O} fraction (panels \textbf{(a)} to \textbf{(c)} respectively) for an $r=\qty{2}{nm}$ nanopore with 720 \ce{H2O} molecules.} \label{fig: VDOS PPR f_H2O}
\end{figure}

Figure \ref{fig: VDOS PPR f_H2O} shows a comparison between the VDOS, the phonon participation ratio, and the \ce{H2O} fraction. The PPR shows that the intercept is primarily due to localized modes. From the \ce{H2O} fraction, we identified those localized modes from the \ce{H2O} molecules. Similarly, modes with energy above \qty{200}{meV} are localized. The temperature significantly affects the localization. For example, some localized modes between \qty{50}{meV} and \qty{100}{meV} become mostly delocalized at high temperature. 

\section{Conclusions}
\label{sec: conclusions}

In this study, we performed molecular dynamics simulations of \ce{SiO2} nanopores and characterized their VDOS as a function of force field, crustal structure, and temperature. We also investigated the effects of the addition of surface \ce{H2O} molecules. The simulation results were compared with experimental data from time-of-flight inelastic neutron scattering and phonon calculations from DFT.

The tested force fields capture some features of the experimental PDOS. None reproduces the overall PDOS profile of the $\alpha-$quartz bulk material. Remarkably, two of the force fields (ReaxFF2023 and PoreMS) produce a relatively accurate representation of the amorphous bulk \ce{SiO2}. The force field correlates well with DFT phonon calculations at low energies. Overall, we consider the tested force fields useful for \ce{SiO2} nanopre simulations. 

We found that the VDOS is relatively temperature-independent in the 100K to 300K range. There is a slight softening of phonon modes, which is more pronounced when using the PoreMS force field. This is attributable to a low amount of phonon anharmonicity.

A comparison between the nanopore and the bulk VDOS shows broadening of the modes, particularly for the $\alpha$-quartz and $\beta$-Cristobalite  structures. The amorphous pore shows relatively low broadening. 

The addition of H2O molecules to the nanopore significantly affects the calculated VDOS. The main effect is shown at the low energies, in particular at the intercept with \qty{0}{meV}. The VDOS shows the molecular diffusion, which depends on temperature and loading. In particular, we observe an approximately quadratic scaling between \qty{100}{K} and \qty{200}{K}, expected for a solid, and an almost linear scaling beyond \qty{250}{K}, characteristic of liquids. The RMSD exhibits similar behavior, with the diffusion coefficient increasing markedly near \qty{250}{K}. The ReaxFF2023 force field shows significant reactivity between the pore and the \ce{H2O} molecules. We found that most of the modes at energies larger than \qty{200}{meV} are localized modes. We observed that temperature delocalizes some modes near \qty{100}{meV}. 

Overall, this work advances the understanding of \ce{SiO2} nanopores by systematically characterizing the PDOS under a range of relevant environmental conditions. These results provide a framework that can be extended to other similar nanomaterials. All data, input parameters, and scripts are publicly available \cite{MDTools, GalavizGit2025}, enabling studies of other nanopores with relatively minor modifications. The system studied here is inherently complex and warrants further investigation. In particular, simulating biomolecule capture has important applications. This topic is left for future work.

\section{Acknowledgment}
\label{sec: acknowledgment}

This work was supported in part by JSPS KAKENHI Grant Number 25K01794.
This research was undertaken with the assistance of resources and services from the National Computational Infrastructure (NCI), which is supported by the Australian Government. This work was supported by resources provided by the Pawsey Supercomputing Research Centre’s Setonix Supercomputer, with funding from the Australian Government and the Government of Western Australia. We acknowledge the support of the Australian Government in providing access to the Australian Centre for Neutron Scattering, which is partly funded through the National Collaborative Research Infrastructure Strategy (NCRIS).

\appendix
\section{Annealing test}
\label{app: annealing_test}

\begin{figure}[!ht]
\centering
    \includegraphics[width=1\textwidth]{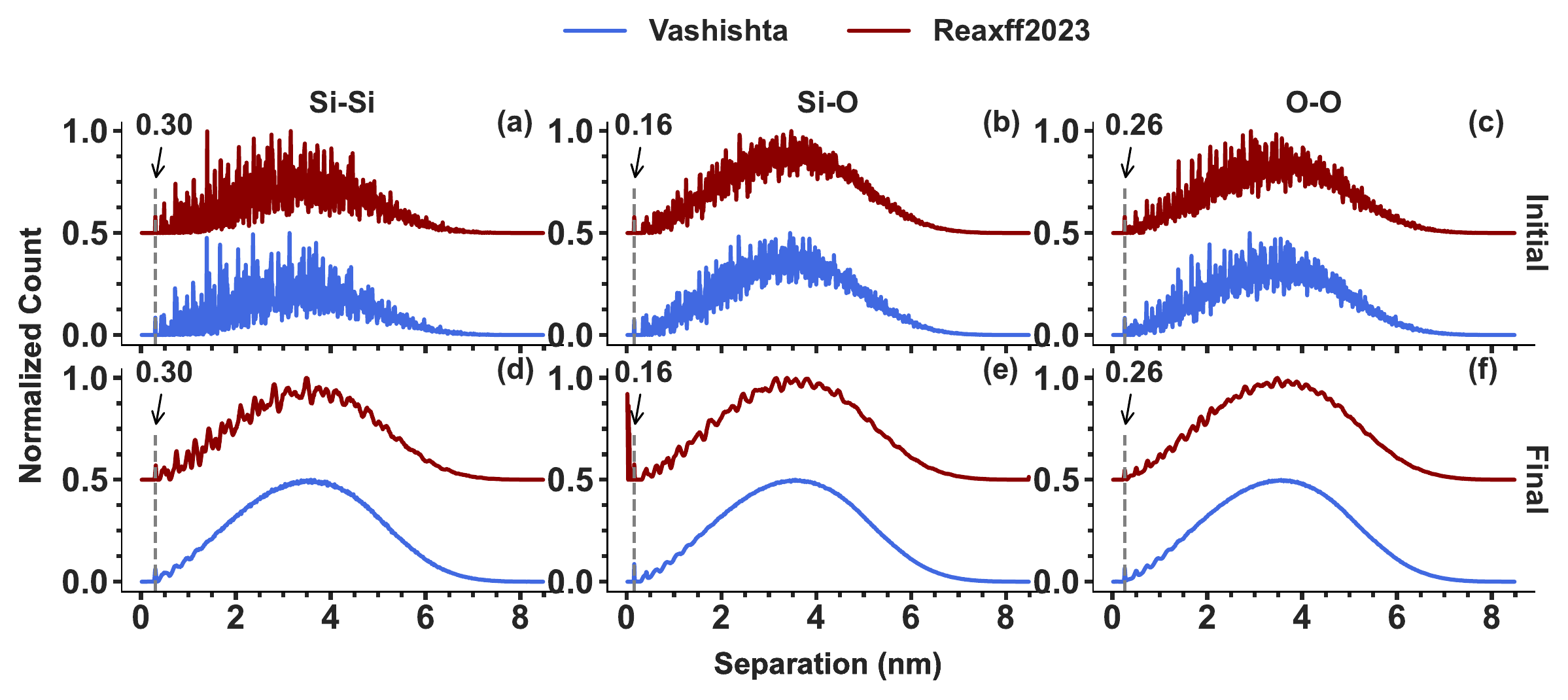}
\caption{Pair distribution histogram. Panels \textbf{(a)} to \textbf{(c)} show the initial pair distribution for each pair: Si-Si, Si-O, and O-O. The vertical dashed line and corresponding arrow indicate the minimal typical separation for each pair. Panels \textbf{(d)} to \textbf{(f)} show the result after simulating the annealing procedure.} \label{fig: annealing_test}
\end{figure}

We made an amorphous structure by using the annealing Protocol \ref{protocol: annealing}. The procedure is illustrated in Figure \ref{fig: annealing}, where we see the initial and final states structures (Figure \ref{fig: annealing}-\textbf{(a)} and \textbf{(c)}, respectively). In addition, we quantified the structure's state by calculating a pair distribution histogram. Figure \ref{fig: annealing_test} shows the result. The top panels show the initial distribution. We observe a discrete distribution characteristic of a crystal structure. The vertical dashed line shows the separation of the first neighbor for a particular pair. Notice that the ReaxFF2023 is shifted vertically by 0.5 units for visualization purposes. The same metric for the final structure shows an almost smooth curve characteristic of an amorphous distribution. Notice that the ReaxFF2023 result shows a significant increase of Si-O pairs with distances closer than the initial \qty{0.16}{nm} separation. 

\section{Heating procedure test}
\label{app: heating_procedure_test}

\begin{figure}[!ht]
\centering
    \includegraphics[width=\textwidth]{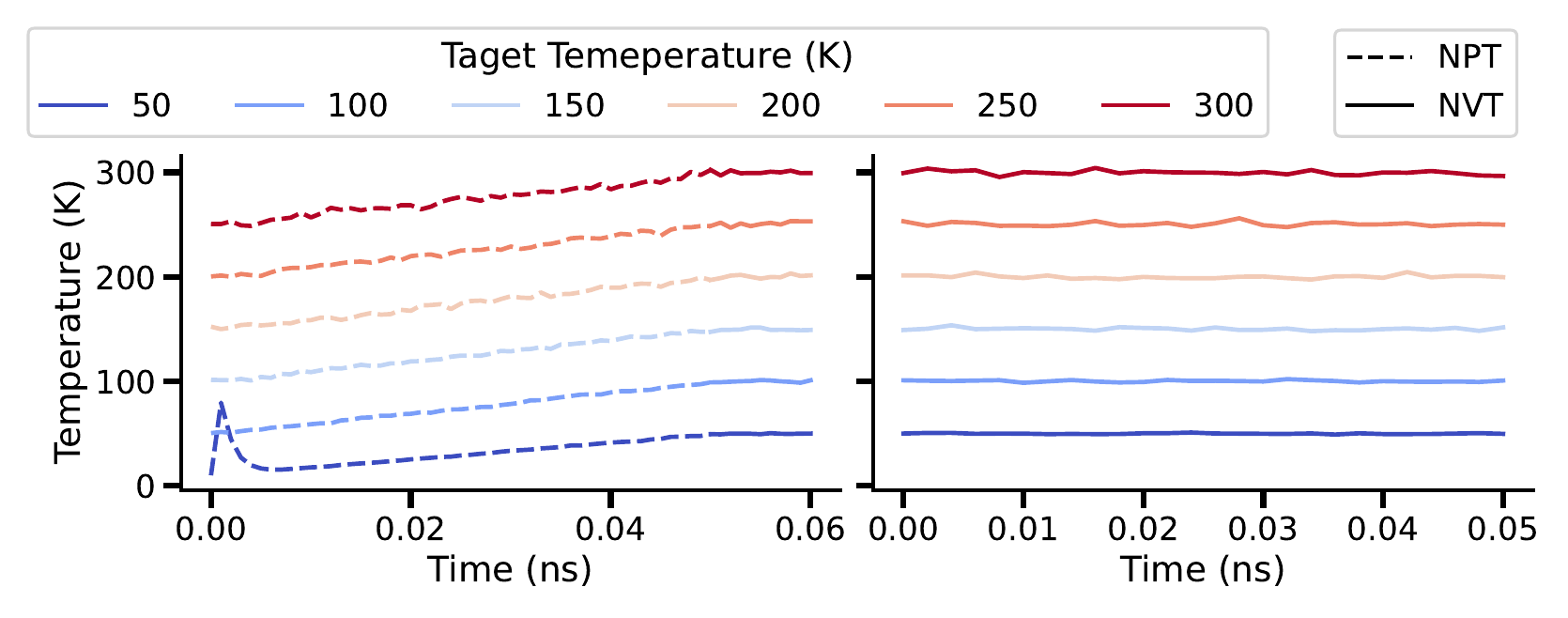}
\caption{Simulation protocol.} \label{fig: protocol2}
\end{figure}

We tested an alternative simulation protocol where the system is progressively heated (see Figure \ref{fig: protocol2}). Besides the initial ramp-up of the temperature, the procedure seems to evolve without deficiencies. 

\begin{figure}[!ht]
\centering
    \includegraphics[width=0.85\textwidth]{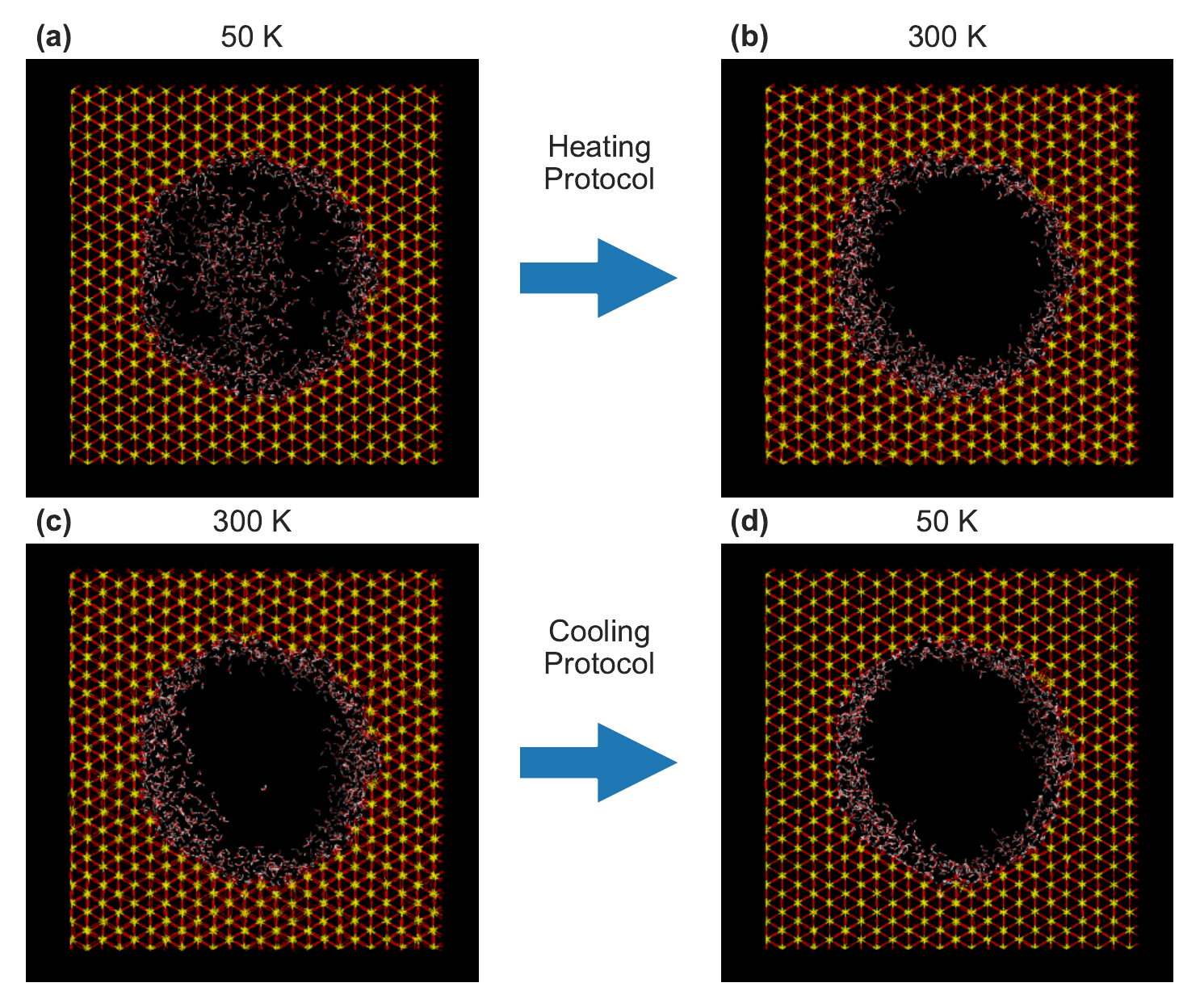}
\caption{Simulation protocol. Panel \textbf{(a)} shows the initial state of the heating protocol. Panel \textbf{(b)} shows the final state. Panels \textbf{(c)} and \textbf{(d)} show the equivalent states for the cooling protocol, respectively.} \label{fig: structure state}
\end{figure}

However, when looking at the structure, we notice that the \ce{H2O} molecules at low temperature are distributed in the center of the nanopore. at the beginning, the \ce{H2O} molecules are randomly distributed inside the pore. Therefore, when starting from a low temperature, the molecules don't diffuse enough to cover the pore's wall. In contrast, if the system is first heated to \qty{300}{K}, the \ce{H2O} molecules can properly distribute around the pore's wall. The two cases are shown in Figure \ref{fig: structure state}. 

\bibliographystyle{elsarticle-num} 
\bibliography{main}







\end{document}